\documentclass[preprint,showkeys,amsmath,amssymb,aps]{revtex4-2}
\usepackage{latexsym}
\usepackage{graphics}
\usepackage{color}
\definecolor{navyblue}{rgb}{0,0,0.8}
\usepackage{graphicx,amssymb,epsfig,epsf,amsmath}
\usepackage{physics}
\usepackage[colorlinks=true,linktocpage=true,allcolors=navyblue]{hyperref}

\def\be{\begin{equation}}
	\def\lan{\left\langle}
	\def\ran{\right\rangle}
	\def\ee{\end{equation}}
\def\barr{\begin{array}}
	\def\earr{\end{array}}

\def\l{\left}
\def\r{\right}
\def\dis{\displaystyle}
\def\ed{\end{document}}

\def\cac{{\cal C}}

\def\can{{\cal N}}

\def\bH{{\bf H}}

\def\kp{{\kappa}}
\begin{document}

\title[Thermalization in many-fermion quantum  systems]{Thermalization in many-fermion quantum systems with\\one- plus random $k$-body interactions }%

\author{Priyanka Rao}
\email{pnrao-apphy@msubaroda.ac.in}
\author{N. D. Chavda}
\thanks{Corresponding author}
\email{ndchavda-apphy@msubaroda.ac.in}
\affiliation{Department of Applied Physics, Faculty of Technology and Engineering, The Maharaja Sayajirao University of Baroda, Vadodara-390001, India}

\begin{abstract}
We study  {the mechanism of} thermalization in finite many-fermion systems with random $k$-body interactions in presence of a mean-field. The system Hamiltonian $H$, for $m$ fermions in $N$ single particle states with $k$-body interactions, is modeled by mean field one-body $h(1)$ and a random $k$-body interaction $V(k)$ with strength $\lambda$. Following the recent application of $q$-Hermite polynomials to these ensembles,
a complete analytical description of parameter $q$, which describes the change in the shape of state density from Gaussian for $q=1$ to semi-circle for $q=0$ and intermediate for $0<q<1$, and variance of the strength function are obtained in terms of model parameters. The latter gives the thermalization marker $\lambda_t$ defining the thermodynamic region. For $\lambda \ge \lambda_t$, the smooth part of the strength functions is very well represented by conditional $q$-normal distribution ($f_{CN}$), which describes the transition in strength functions from Gaussian to semi-circle as the $k$-body interaction changes from $k = 2$ to $m$ in $H$. In the thermodynamic region, ensemble averaged results for the first four moments of the strength functions and inverse participation ratio (IPR) are found to be in good agreement with the corresponding smooth forms. For higher body rank of interaction $k$, system thermalizes faster.
\end{abstract}


\maketitle
%\tableofcontents

\section{Introduction}
 {
The onset of thermalization in isolated many-body quantum systems, due to inter-particle interaction between their constituents, has received considerable interest since quite a long time and is now an important research area \cite{Znid08,Pal-10,Khatami2012,JHB2012,nandkishore-rev,Huse2007,Altman2015}. The connection between the mechanism that leads to thermalization in many-body quantum systems and the onset of quantum chaos, when inter-particle interactions are strong enough, is well established~\cite{Sengupta,Rigol16,BISZ2016}. In 1995 using ultra-cold atomic gases, Bose-Einstein condensation (BEC) was achieved for the first time in laboratory \cite{BEC-1,BEC-2,BEC-3}. Subsequently experiments on cooling and trapping of ultracold atomic gases \cite{Jorden-08,Schneider-08} and electrons in solids\cite{Perfetti-06} developed tremendously, along with important numerical algorithms and techniques to study the dynamics of these isolated quantum systems. This experimental progress facilitates one to artificially simulate these isolated systems and also control the interaction strength by which transition between different quantum phases can be studied \cite{rigol-2010}. Now, it is clear that the lack of thermalization in many-body interacting systems is due to the emergence of extensively many local integrals of motion, a phenomenon called many-body localization. While on the other hand, the thermalization in non-integrable quantum systems occurs due to the underlying mechanism called the eigenstate thermalization hypothesis (ETH)~\cite{Rigol-2}. ETH essentially means the condition where the individual eigenstates of a many-body interacting system can be treated as (quasi)random  superposition of a large number of the basis states~\cite{Deutsch-94,Rigol-1,San-10,Flam-97,lts-2017} and then the expectation values of typical observables in the eigenstates follow ergodicity principle. Therefore, the study of mechanism of thermalization in terms of properties of chaotic eigenstates has also received considerable importance \cite{BISZ2016,San-10, Relano-2010,PLA-2021} rather than the thermalization itself which is defined according to quantities such as temperature, thermodynamic entropy, or/and time relaxation to the steady-state distribution. Many groups have addressed the problem of the emergence of thermalization in many-body quantum systems using different models and studied various aspects
related to the role of localization and chaos, statistical relaxation, eigenstate thermalization and ergodicity principle \cite{Rigol-1,Rigol-2,Rigol-3,BISZ2016,nandkishore-rev,San-12a,San-12b,kc-18b,kota-therm,Ch-PLA}.
}

 {
In the past, thermalization in fermion and boson systems have been studied in detail using two-body ($k=2$) random embedded Gaussian orthogonal ensemble in presence of a mean field [denoted as EGOE(1+2)] \cite{Flam-97,Flam-96,kota-book,An-04,Ma-10,kota-therm,Ch-PLA} as they form generic models for finite isolated interacting many-particle (fermions or bosons) systems. Going beyond this, the higher body interactions ($k>2$) are found to play an important role in various fields like nuclear physics \cite{Launey}, strongly interacting quantum systems \cite{Blatt,Hammer}, in quantum transport in disordered networks connected by many-body interactions \cite{centro-1,centro-2,Ortega2018} and also in quantum black holes \cite{Cotler-2017,Garcia-2018} and wormholes \cite{Garcia-2019} with Sachdev–Ye–Kitaev (SYK) model. This leads to the recent enhance of interest in random matrix ensembles in many-particle spaces generated by random $k$-body ($k>2$) interactions in presence of a mean-field recently and they are denoted by EGOE(1+$k$) \cite{Manan-Ko,KM2020,KM2020c,PLA-2021,ndc}. In the strong interaction domain the EGOE(1+$k$) reduces to EGOE($k$) and these ensembles generate Gaussian to semi-circle transition in the eigenvalue density as body rank $k$ of the interaction changes from 1 to $m$~\cite{MF75,brody81}.
}

 {
Recently, the embedded ensembles with higher $k$-body interactions are analyze by employing $q$-Hermite polynomials \cite{Manan-Ko,KM2020,KM2020c,PLA-2021,ndc}. The use of $q$-Hermite polynomials in embedded ensembles appears due to the recent work by Verbaarschot et al on the study of spectral density using SYK models, which are embedded ensemble models of Majorana fermions \cite{Cotler-2017,Garcia-16,Garcia-17}. It has been demonstrated that the $q$-normal distribution $f_{N}$, which is the weight function for $q$-Hermite polynomials \cite{SZAB}, can be used to describe the smooth form of spectral density in EGOE($k$) \cite{Manan-Ko,PLA-2021}. Further, the formulae for $q$-parameter in terms of model parameters $N$, $m$ and $k$ are derived for fermion as well as boson systems for both EGOE($k$) and their Unitary variants EGUE($k$) \cite{Manan-Ko}. Very recently, the condition of thermalization is studied in many-body bosonic systems with $k$-body interaction modeled by EGOE(1+$k$) in terms of properties of chaotic wave-function and the formula for thermalization marker $\lambda_t$ in terms of model parameters is derived \cite{PLA-2021}. Also, it is shown that \cite{PLA-2021} in the strong interaction domain ($\lambda >> \lambda_t$), the strength functions (also known as local density of states) make transition from Gaussian to semi-circle as the rank of interaction $k$ increases in EGOE(1+$k$) and their smooth forms can be represented by the conditional $q$-normal distribution function $f_{CN}$ to describe this crossover. Moreover, the interpolating form of strength function given by $f_{CN}$ is used to describe the fidelity decay after random $k$-body interaction quench and the formula for NPC valid in thermalization region is derived \cite{PLA-2021}. It is important to note that for bosonic systems, in addition to dilute limit (defined by $m \rightarrow \infty$ , $N \rightarrow \infty$, and $m/N \rightarrow 0$), the dense limit  (defined by $m \rightarrow \infty$ , $N \rightarrow \infty$, and $m/N \rightarrow \infty$) is also feasible. For fermion systems, only dilute limit is possible. Therefore the focus was on the dense limit in the analysis of bosonic systems. Very recently, for fermionic EGOE(1+$k$), the wave-function structure is analyzed by deriving the lower order moments of the strength functions in $m$-fermion spaces \cite{KM2020c}.
}

 {
Going further, the purpose of the present study is to understand the mechanism of thermalization, in terms of the structure of chaotic eigenfunctions of isolated many-fermion systems, by varying the strength parameter $\lambda$ of the $k$-body interaction in the EGOE(1+$k$) Hamiltonian given by $H=h(1)+\lambda V(k)$, where $h(1)$ is one-body part of the interaction defined by single particle (sp) energies and $V(k)$ is the random $k$-body interaction. The focus is in the dilute limit as dense limit is not feasible for fermionic systems. The strength functions and measures based on them are basis dependent, we employ mean-field defined by fixed one-body part, as was done in the previous studies \cite{Manan-Ko,KM2020c,PLA-2021}. We utilize $q$-normal distribution $f_N$ and conditional $q$-normal distribution $f_{CN}$ to describe the smooth forms of state density as a function of $\lambda$ and strength function in strong interaction domain, respectively. The form of these functions can be completely determined by the parameters, the correlation coefficient $\zeta$ and $q$. A complete analytical description of $\zeta$ and $q$ as a function of ($m,N,k$) are derived and they are tested with the ensemble averaged numerical results. Also, we analyze the lower moments of the so-called strength functions (also known as local density of states) by varying interaction strength $\lambda$ and obtain a formula giving the critical strength $\lambda=\lambda_t$, in terms of model parameters $m$, $N$ and $k$, for the onset of thermalization rather than choosing some specific value $\lambda$ such that $\lambda>>\lambda_t$~\cite{Manan-Ko,KM2020c}. Moreover, the region of thermalization is identified by comparing  the moments of $f_{CN}$ with that of the strength functions and shown that for $\lambda \ge \lambda_t$, the match between them is very good giving the region beyond which the strength functions follow $f_{CN}$.  Further, the method of construction of strength functions is described so that their smooth forms can be correctly represented by $f_{CN}$. Results for inverse participation ratio are compared with smooth forms and found good agreement in the thermalization region.
}

The paper is organized as follows: The embedded ensembles EGOE(1+$k$) are briefly defined in Section~\ref{model}.  Section~\ref{q-hermite} gives  preliminary description about $q$-Hermite polynomials and conditional $q$-normal distributions. Results of state density along with a complete analytical description of the correlation coefficient $\zeta$ and $q$-parameter for fermionic EGOE(1+$k$) as a function of $\lambda$ and body rank $k$  are given in Section \ref{sd}. In Section \ref{SF}, the method of construction of strength functions is described for EGOE(1+$k$) giving Gaussian to semi-circle transition as body rank $k$ increases from 2 to $m$ in the strong interaction domain. More importantly, it is demonstrated that the conditional $q$-normal distribution function $f_{CN}$ describes this transition correctly. Further, the first four moments of strength functions are studied and using the width of these strength functions, analytical formula for the chaos marker $\lambda_t$ is derived in Section \ref{zeta}. Results for our calculations for IPR are presented in Section~\ref{NPC}. Finally, Section \ref{conclusions} gives the concluding remarks.

\section{Ensembles for fermion systems - EGOE(1+$k$)}
\label{model}

 {Interactions between the constituents of finite isolated quantum systems, such as atomic nuclei, complex atoms and molecules, mesoscopic devices like small metallic grains and quantum dots, spin systems modeling a quantum computing core and so on, are known to be few-body in character. The random matrix ensembles (Gaussian orthogonal ensembles (GOE)) which incorporate this feature along with nature of constituents (fermions, bosons, Majorana fermions and so on) are called embedded ensembles (Embedded Gaussian orthogonal Ensembles (EGOE)). Now we will briefly describe these ensembles for fermion systems.}
	
 {Consider a system consisting of $m$ spinless fermions occupying $N$ sp states and interacting with each other via $k$-body ($k \leq m$) interactions. Distributing $m$ such fermions in $N$ degenerate sp states generates $m$-particle Hilbert space of dimension $d={N \choose {m}}$. With sp states denoted by $n_i$, a basis state in $k$-particle space, in occupation number representation, is denoted by $|n_1 n_2 \cdots n_k \rangle$ with $\sum_{i=1}^{k} n_i =k$ and in increasing order, $n_1 < n_2 < \cdots < n_k$. $m$-particle basis states can be defined similarly. The $k$-body random Hamiltonian $V(k)$ for such a system in second quantization form can be defined as,
\be
V(k) = \dis\sum_{\alpha,\beta} w_{k;\alpha,\beta}\, F^\dag(k;\alpha) F(k;\beta)\;.\label{eq.egoek}
\ee
Here, $\alpha$ and $\beta$ are the $k$-particle states in occupation number representation. $V(k)$ generates a $k$-particle Hilbert space of dimension $d_k={N \choose k}$ with anti-symmetrized matrix elements $w_{k;\alpha,\beta}$ which are independent random variables. As the $V(k)$ matrix is chosen to be GOE in $k$-particle space, the $k$-particle matrix elements $w_{k;\alpha,\beta}$ are Gaussian random variables with zero mean and unit variance. The variance of diagonal matrix elements is $\overline{w^2_{k;\alpha,\alpha}}=2$ while that of off-diagonal matrix elements, for ($\alpha \ne \beta$), is $\overline{w^2_{k;\alpha,\beta}}=1$. Further, $F^{\dag}(k;\alpha) = \prod_{i=1}^{k} f_{n_i}^{\dag}$ and $F(k;\alpha) = (F^{\dag}(k;\alpha))^{\dag}$ with $F^{\dag}(k;\alpha)$ (and $F(k;\alpha)$) are creation (and destruction) operators respectively for $k$-particle fermion state $\alpha$. Similarly, $f_{n_i}^{\dag}$ (and $f_{n_i}$) are the fermion creation (and destruction) operators respectively for sp state $n_i$. The action of this $V(k)$ on the $m$-particle basis states generates EGOE($k$). Due to the $k$-body interactions, many of the matrix elements of the $m$-particle Hamiltonian matrix of EGOE($k$) are zero for $m>k$ while many of the nonzero matrix elements are linear combinations of $k$-body matrix elements. This is different from what we have in the case of GOE. However when $k=m$, EGOE($k=m$) is identical to GOE by construction. Detailed study on these ensembles can be found out in \cite{kota-book} and references therein.}

 {Now, adding the mean-field part generated by the presence of other fermions in the system, one can have EGOE($1+k$) defined by,
\be
H= h(1) + \lambda V(k)\label{eq.egoe1+k}
\ee
Here, the mean-field given by the one-body operator $h(1)=\sum_{i=1}^N \varepsilon_i f^{\dag}_{n_i} f_{n_i}$ is described by sp energies $\varepsilon_i$ with average spacing $\Delta$; $f^{\dag}_{n_i} f_{n_i}$ is the number operator for the $i$th sp state. Throughout this work, we use fixed sp energies $\varepsilon_i = i + 1/i$ to define $h(1)$, just as in many previous studies \cite{Flam-96a,PLA-2021,Ma-10,CK2017}. Note that the second term $1/i$ in $i$ has been added to avoid the degeneracy of many-particle states as discussed first in \cite{Flam-96a}. The parameter $\lambda$ represents $k$-body interaction strength. With $\Delta=1$, $\lambda$ is measured in units of the average spacing of the sp energies defining $h(1)$.}

 {In the present study, we utilize $q$-normal distribution $f_{N}$ and conditional distribution $f_{CN}$ and the results are presented for the onset of thermalization in these systems using the lower order moments of the strength functions. The numerical results are obtained by fully diagonalizing EGOE(1+$k$) ensemble for various values of body rank $k$ and interaction strength $\lambda$. The following two examples of EGOE(1+$k$) are considered: (i) 100 member ensemble with $m=6$ fermions in $N=12$ sp states and dimensionality of system is $d = 924$; (ii) 20 member ensemble with $m=7$ fermions in $N=14$ sp states and dimensionality of the system is $d=3432$. The ensemble average is carried out by making the spectra of each member of the ensemble zero centered and scaled to unit width.}

\section{$q$-Hermite polynomials and conditional $q$-normal distribution}\label{q-hermite}
L. J. Rogers first introduced $q$-Hermite polynomials in Mathematics to prove the Rogers–Ramanujan identities \cite{ISV-87} and their important properties are due to Szego and Carlitz \cite{szab-2013}. The $q$-Hermite polynomials are related with the Chebyshev, Rogers-Szego, Al-Salam-Chihara polynomials and others. Recently they have found applications in non-commutative probability, quantum physics, combinatorics and so on. Also the $q$-calculus has attracted many researchers working in the field of special functions as it is a very powerful tool in quantum computation \cite{qh-2021}.

Let us begin with $q$-numbers, $\l[n\r]_q = (1-q)^{-1}(1-q^n)$ and $q$-factorials, $[n]_q!=\Pi^{n}_{j=1} [j]_q$, which define the $q$-Hermite polynomials. We have $[n]_{q \rightarrow 1}=n$. Also, $[0]_q!=1$.
The recursion relation that defines the $q$-Hermite polynomials $\bH_n(x|q)$ is given as \cite{ISV-87},
\be
x\,\bH_n(x|q) = \bH_{n+1}(x|q) + \l[n\r]_q\,\bH_{n-1}(x|q).
\ee
Here, $\bH_0(x|q)=1$ and $\bH_{-1}(x|q)=0$. The $q$-Hermite polynomials with $q=1$ give normal Hermite polynomials, which are related to Gaussian and with $q=0$ give Chebyshev polynomials, which are related to semi-circle. Importantly, the weight function of $q$-Hermite polynomials is the $q$-normal distribution $f_{N}(x|q)$ defined by \cite{KM2020},
\be
f_{N}(x|q)  =  \dis\frac{1}{2 \pi}\,\sqrt{\frac{1-q}{4-(1-q)x^2}} \dis\prod_{i = 0}^{\infty} [(1+q^i)^2-(1-q)q^i x^2]\,(1-q^{i+1}) \,.
\label{eq.ent7}
\ee
$q$-Hermite polynomials are orthogonal within the limits $s(q)=[-2/\sqrt{1-q}, 2/\sqrt{1-q}]$, which can be inferred from the following property,
\begin{equation}
	\int_{s(q)}  dx \,\bH_n(x|q) \,\bH_m(x|q) f_{N}(x|q)\;= [n]_q! \, \delta_{mn}.
	%\label{eq:zfcqn}
\end{equation}

Here, $x$ is normalized variable, with  zero centroid and unit variance, and $-2/\sqrt{1-q} \leq x \leq 2/\sqrt{1-q}$. In general $q$ can take any value from [-1,1], however in the present study $q \in [0,1]$. Note that $\int_{s(q)} dx\, f_{N}(x|q)=1$ and one can see that in the limit $q \rightarrow 1$, $f_{N}(x|q)$ will take Gaussian form
and in the limit $q \rightarrow 0$, $f_{N}(x|q)$ will take semi-circle form.

Further, the bivariate $q$-normal distribution $f_{biv-qN}(x, y|\zeta, q)$ with standardized variables $x$ and $y$ is defined as \cite{SZAB,KM2020},

\begin{equation}
%	\barr{rcl}
	f_{biv-qN}(x, y|\zeta, q) = f_{N}(x|q)\, f_{CN}(y|x;\zeta,q)\\\\
	=f_{N}(y|q)\, f_{CN}(x|y;\zeta,q).
%	\earr
	\label{eq:bivq}
\end{equation}
Here $\zeta$ is the bivariate correlation coefficient. {$f_{N}(x|q)$ and $f_{N}(y|q)$ are the marginal densities of $f_{biv-qN}(x, y|\zeta, q)$}. Then, the conditional $q$-normal densities, $f_{CN}$ can be given as,
\begin{equation}
	\barr{rcl}
	f_{CN}(x|y;\zeta,q) &=& f_{N}(x|q) \; v(x,y|\zeta,q);\\\\
	
	f_{CN}(y|x;\zeta,q) &=& f_{N}(y|q)\; v(x,y|\zeta,q);\\\\
	
	v(x,y|\zeta,q)&=&
	\dis\prod_{i=0}^{\infty} \frac{(1-\zeta^2\, q^i)}{(1-\zeta^2 q^{2i})^2-(1-q)\zeta q^i (1+\zeta^2\,q^{2i}) \,x\,y + (1-q)\,\zeta^2\,(x^2 + y^2)\,q^{2i}}\;.
	\earr
	\label{eq:biv-cqn}
\end{equation}
Also,
\begin{equation}
	\int_{s(q)} dx \, \bH_n(x|q)\, f_{CN}(x|y;\zeta,q) = \zeta^n\, \bH_n(y|q).
\end{equation}
From the above relation, it is clear that $f_{CN}$ and $f_{biv-qN}$ are normalized to 1 over the range $s(q)$.

 {Recently, it has been shown that the smooth forms of eigenvalue density and strength functions for embedded ensembles with $k$-body interactions, EGOE($k$), can be described $f_{N}$ (equation~\ref{eq.ent7}) \cite{Manan-Ko,kc-18b} and by $f_{CN}$ (equation~\ref{eq:biv-cqn}) \cite{PLA-2021} respectively giving transition from Gaussian form to semi-circle form as $k$ changes from $2$ to $m$. Going beyond this, we will show in the nest section that $f_{N}$ can also describe the shape of the state density (intermediate form between Gaussian to semi-circle) as the interaction  $\lambda$ increases in fermionic EGOE(1+$k$) just like bosonic systems~\cite{PLA-2021}.  Also, given is a complete analytical description of variation of  {correlation coefficient $\zeta$ and} parameter $q$ with $\lambda$.}

 \section{State density and formula of correlation coefficient $\zeta$ and parameter $q$ for EGOE(1+$k$)}
\label{sd}

\subsection{State density}
\begin{figure}[tbh!]
	\centering
	\begin{tabular}{ll}
		\includegraphics[width=0.5\textwidth]{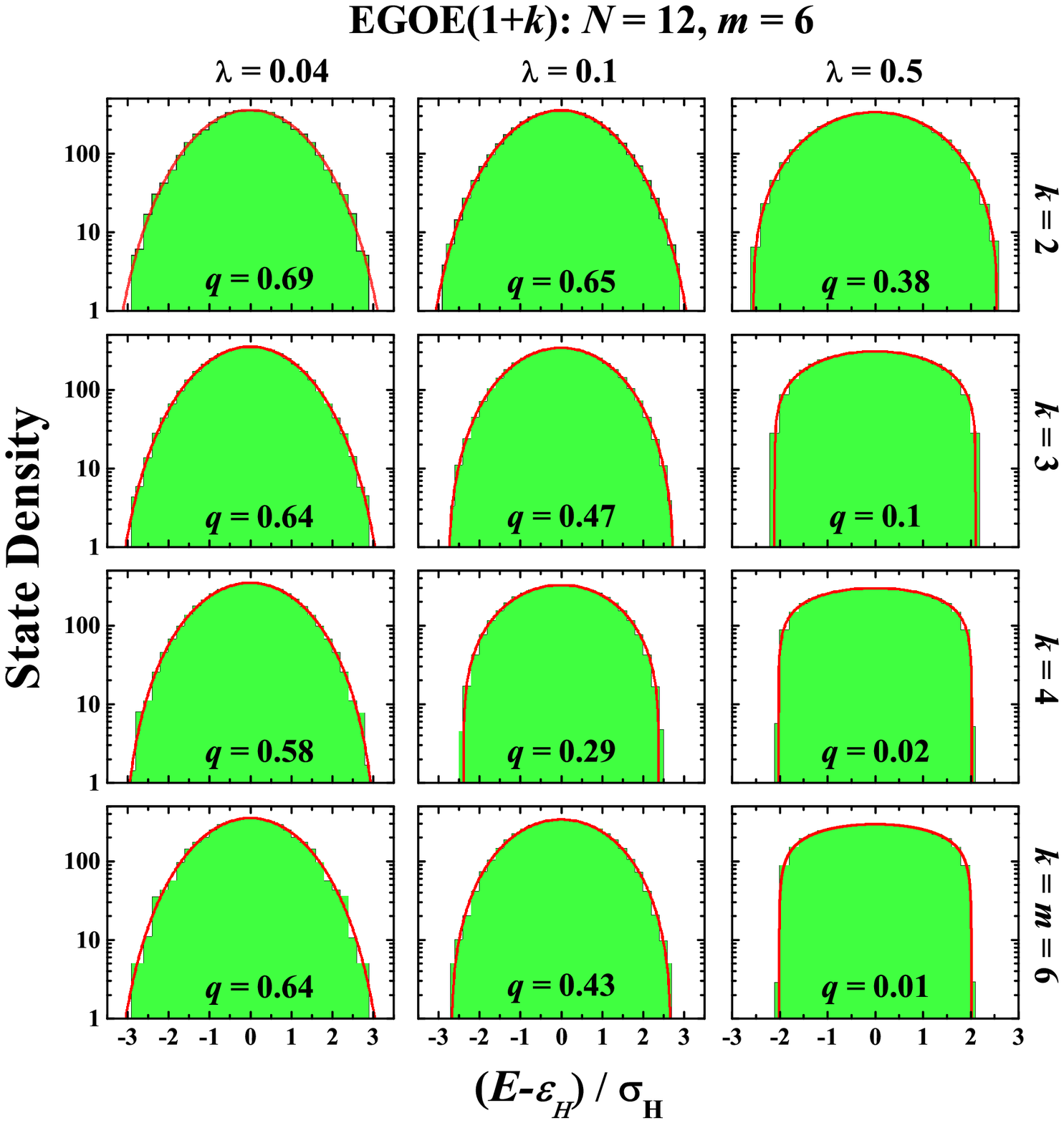}&\includegraphics[width=0.5\textwidth]{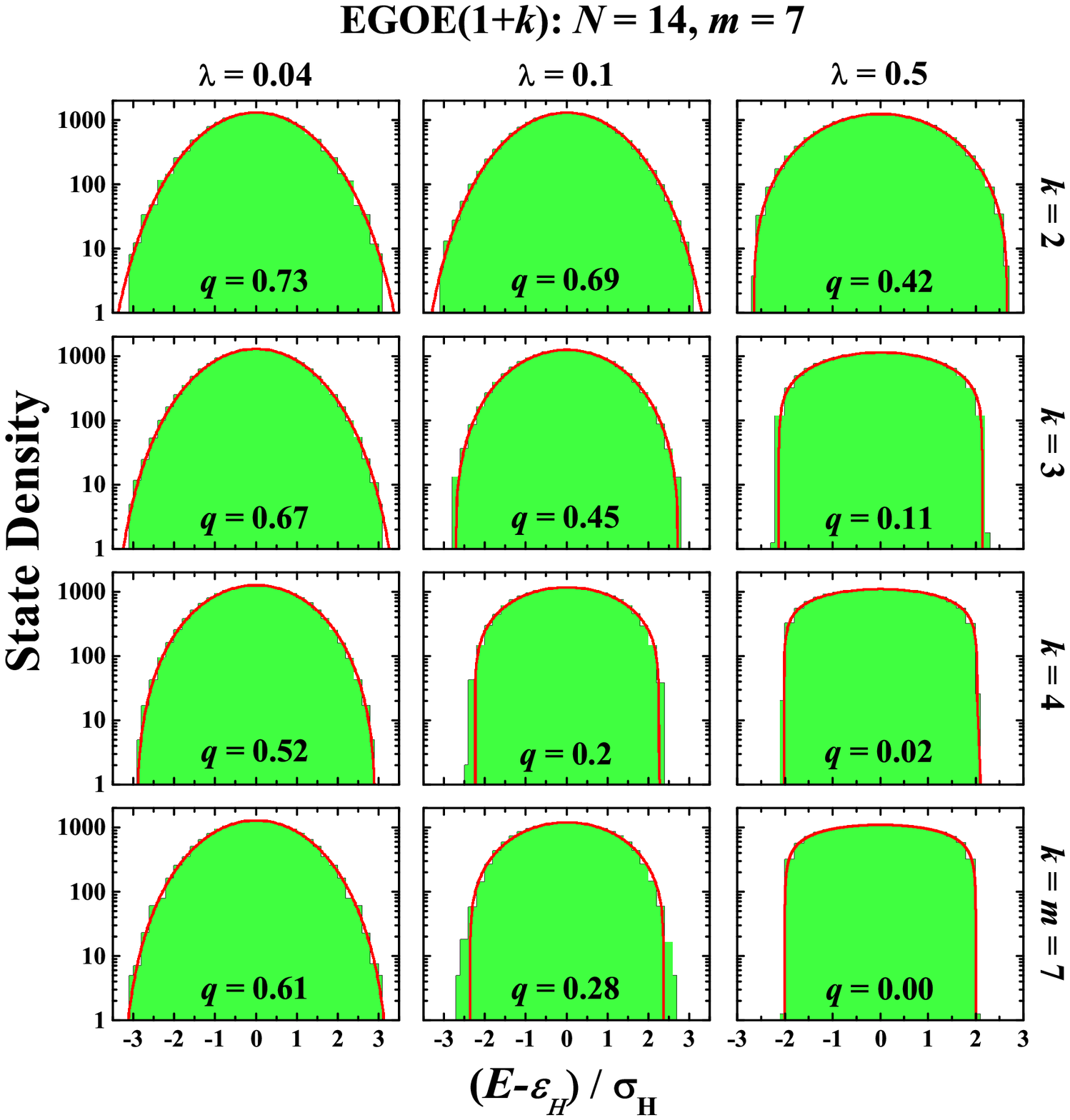}\\
		(a)&(b)
	\end{tabular}
	\caption{Ensemble averaged state density $\rho(E)$ vs. normalized energy $E$ results for EGOE($1+k$) examples  (a) a 100 member ensemble with $m = 6$ fermions in $N=12$ sp states and (b) a 20 member ensemble with $m = 7$ fermions in $N=14$ sp states. Results are shown for different values of interaction strength $\lambda$ = 0.04, 0.1, and 0.5 for $k = 2$, 3, 4 and $k = m$. Ensemble averaged state density histograms are compared with   {$\rho(E)$ given by equation~\ref{eq:fqne}} (red smooth curves) with the corresponding ensemble averaged $q$ values given in each panel. In all the plots $\int dE  \rho(E) = d$.}\label{fig.rho-q-f-1}
\end{figure}

In figure~\ref{fig.rho-q-f-1} histograms represent ensemble averaged state density vs. normalized energy for EGOE(1+$k$) examples for various values of $k$ and for different interaction strength $\lambda$. Results are shown for EGOE(1+$k$) ensemble with $m = 6$ fermions occupying $N = 12$ sp states (figure~\ref{fig.rho-q-f-1}(a)) and with $m = 7$ fermions occupying $N = 14$ sp states (figure~\ref{fig.rho-q-f-1}(b)). The sp energies defining the one body Hamiltonian $h(1)$ are chosen to be $\varepsilon_i= i + 1/i$, $i = 1, 2, \ldots N$. In the calculation, firstly the spectra of each member of ensemble is zero centered (with centroid $\epsilon_H$) and scaled to unit width (with width $\sigma_H$) and then ensemble averaged histograms are obtained. The smooth red curves in figure~\ref{fig.rho-q-f-1} are obtained using equation~\ref{eq.ent7} and are given as,
\be
\begin{array}{rcl}
	\rho(E)\,dE &=&  \dis dE\,  \frac{\can}{ \sigma_H}\,\sqrt{\frac{1-q}{1-\l( \frac{E-\epsilon_H}{E_0}\r)^2}}\\ && \times \dis \prod_{i = 0}^{\infty} (1-q^{i+1}) \l[(1+q^i)^2-q^i\;4\;\l( \frac{E-\epsilon_H}{E_0}\r)^2 \r];\\\\
	
	&&\dis E_0^2=\frac{4\sigma_H^2}{1-q},\;\; \dis \epsilon_H-\frac{2\,\sigma_H}{\sqrt{1-q}} \leq E \leq \epsilon_H + \frac{2\,\sigma_H}{\sqrt{1-q}}.
	
\end{array}
\label{eq:fqne}
\ee
 {Here, $\can$ is the normalization factor and ensemble averaged values of parameter $q$ are also given in the figure. As the tails of the distribution are an important aspect from a random matrix theory point of view, semi-log scale is used in figure~\ref{fig.rho-q-f-1} in order to see the comparison between the tails of the histograms with the analytical prediction given by equation~\ref{eq:fqne}.} The results clearly show that with small deviations near tails, the $q$-normal distribution $f_{N}$ correctly describes the intermediate form (i.e. between Gaussian to semi-circle) for the state density as $\lambda$ and $k$ changes in EGOE(1+$k$).

%%%%%%%%%%%%%%%%

\subsection{Correlation coefficient $\zeta$ and its variation with $\lambda$} \label{sec-zeta}
\begin{figure}[!ht]
	\centering
	\begin{tabular}{ll}
		\includegraphics[width=0.5\textwidth]{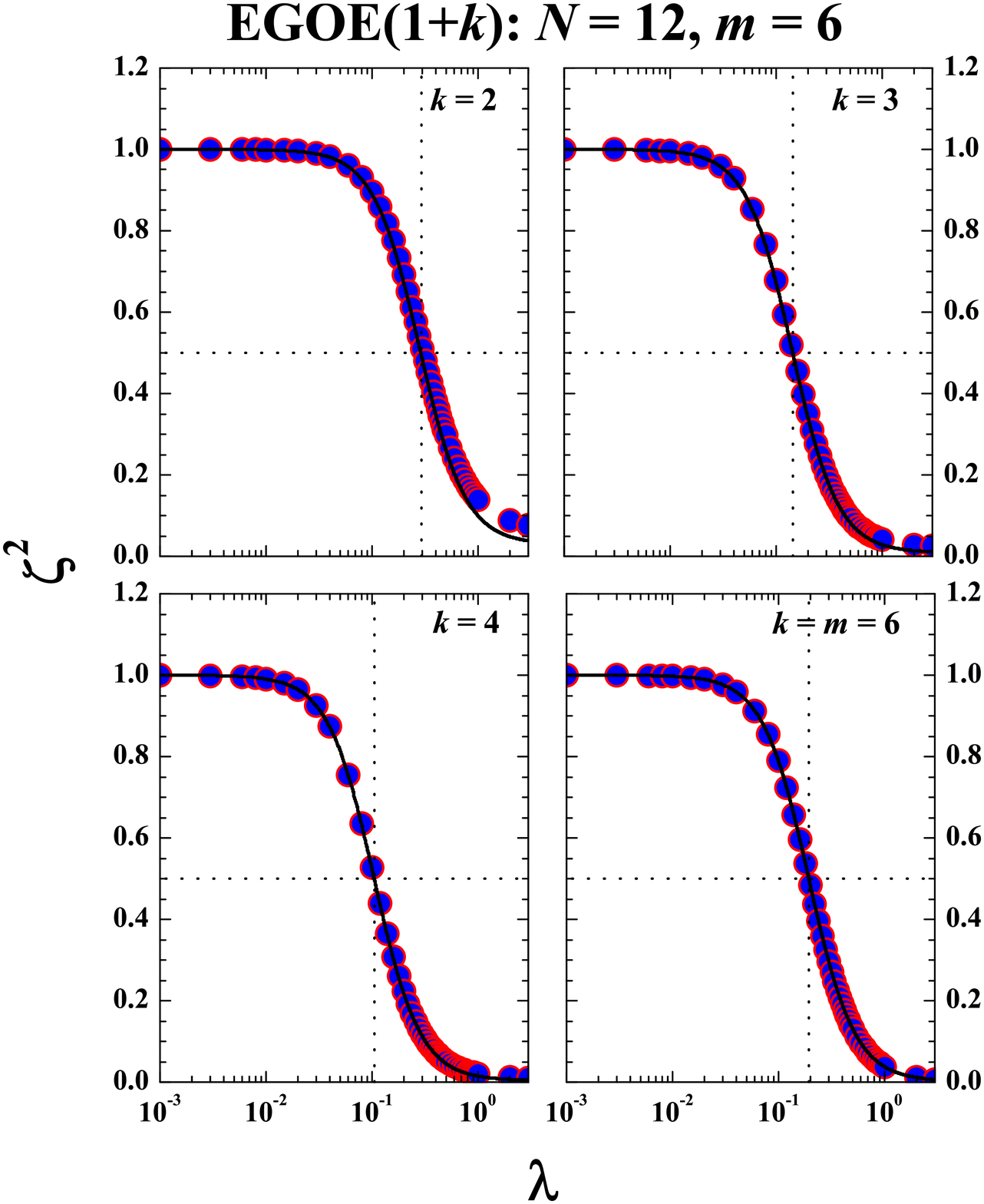}&
		\includegraphics[width=0.5\textwidth]{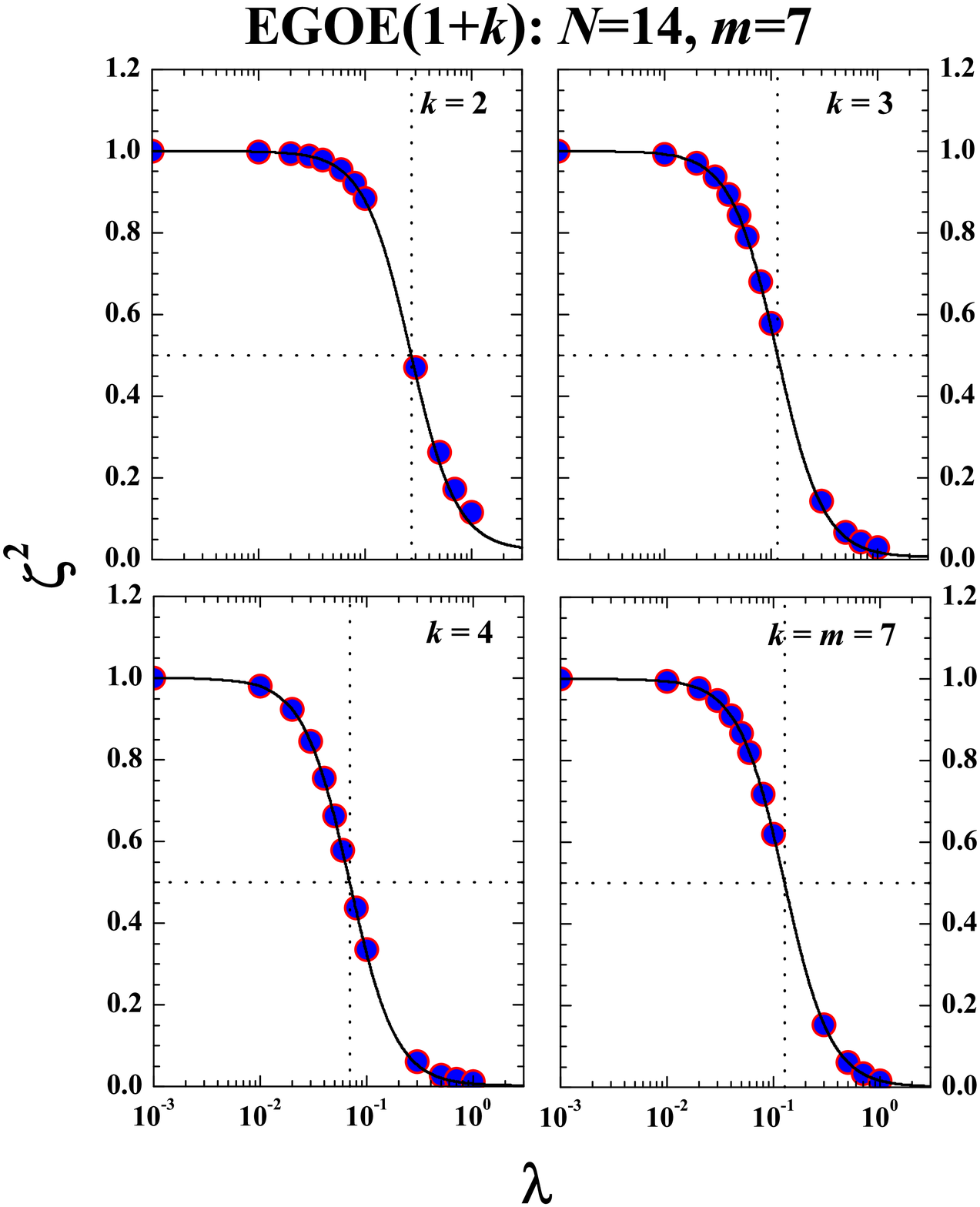}\\
		(a)&(b)
	\end{tabular}
	\caption{Ensemble averaged $\zeta^2$ (solid circles) as a function of interaction strength $\lambda$ for (a) 100 member EGOE(1+$k$) ensemble with $m = 6$ and $N = 12$  and (b) 20 member EGOE(1+$k$) ensemble with $m = 7$ and $N = 14$ for $k$ = 2, 3, 4 and $k = m$. The ensemble averaged results are compared with the analytical curves given by equation~\ref{eq.zeta2-f} (smooth black curves). A very good agreement between ensemble averaged results and analytical formula, given by equation~\ref{eq.zeta2-f}, can be seen for both the examples. The vertical dotted line, in each panel, corresponds to $\lambda_t$ (defined ahead in section~\ref{section5b}), obtained with the condition $\zeta^2=0.5$, giving  thermodynamic region. See text for further details.}
	\label{fig-zeta2-f}	
\end{figure}

 {
The correlation coefficient $\zeta$ between $h(1)$ and full Hamiltonian $H$ is defined as
\be
\zeta =\sqrt{\frac{\sigma^2_{h(1)}}{\sigma^2_{h(1)}+\lambda^2 \, \sigma^2_{V(k)}}}\,.\nonumber
\ee
$\zeta$ is also related to the width of the strength functions; see section~\ref{section5b} ahead for further details. The analytical expression of $\zeta$ in terms of model parameters for fermionic EGOE(1+$k$) can be derived as follows: For $H=V(k)$ i.e. with all sp energies as degenerate, we have \cite{kota-book},
\be
\barr{rcl}
\sigma_{V(k)}^2 &=& \dis P(N,m,k) \binom{N}{k}^{-1} \;  \sum_{\alpha,\beta} \overline{ w^2_{k;\alpha,\beta}}\;, \\\\
\text{where}\;P(N,m,k)& =& \dis  {\binom{N}{k}}^{-1}\,\Lambda^0(N,m,k). %\; \;\;
%\Lambda^0(N,m,k)=\binom{m}{k} \binom{N-m+k}{k}.
\earr
\label{eq.sigvk-f}
\ee
Here, $\alpha$ and $\beta$ denote $k$-particle states. As the $V(k)$ matrix is GOE in $k$-particle space, for EGOE($k$), we have
\be
\sigma_{V(k)}^2 =  P(N,m,k) \left \{ \binom{N}{k} + 1 \right \}.
\ee
For one body $h(1)$ part of the Hamiltonian defined by the external sp energies $\varepsilon_i$ we have,
\be
\barr{rcl}
\sigma_{h(1)}^2 &=& \frac{m(N-m)}{N(N-1)} \; X.
\earr
\ee
Here, $X=\sum \tilde{\varepsilon_i}^2$;  $\tilde{\varepsilon_i}$ are the trace-less sp energies of $i$'th state. With inclusion of the contribution from the diagonal part of $V(k)$, the analytical expression for $\zeta^2$ can be given by,
\be
\zeta^2=\frac{\frac{m(N-m)}{N(N-1)}\; X + 2\; \lambda^2\; P(N,m,k)} {\frac{m(N-m)}{N(N-1)}\; X + \lambda^2\; P(N,m,k)\; \big\{1 + \binom{N}{k}\big\} }\;.
\label{eq.zeta2-f}
\ee
Now we test equation~\ref{eq.zeta2-f} with numerical ensemble averaged EGOE(1+$k$) results. We consider two examples for this analysis: (i) a 100 member EGOE(1+$k$) ensemble with $m$ = 6 fermions in $N$ = 12 sp states and (ii) a 20 member EGOE(1+$k$) ensemble with $m$ = 7 fermions in $N$ = 14 sp states. The variation of $\zeta^2$ as a function of $\lambda$ is obtained for various values of the body rank $k$  for both these examples and results are presented in figure~\ref{fig-zeta2-f}. In all these plots the solid circles represent the numerical ensemble averaged results while the black smooth curves are due to analytical results obtained using equation~\ref{eq.zeta2-f}. It is clear from the figure that  $\zeta^2$ is close to 1 for smaller $\lambda$ and as $\lambda$ increases, $\zeta^2$ goes on decreasing smoothly and approaches 0 for all body rank $k$. The numerical results are in very good agreement with the analytical expression given by equation~\ref{eq.zeta2-f}  for all $k$ values for both the examples. Small discrepancy is due to neglect of induced sp energies given by $V(k)$ for smaller $k$. With increasing $k$, contribution of the induced sp energies will steadily tend to reduce, while that of external sp energies remain fixed.
}

\subsection{Variation of parameter $q$ with $\lambda$}
\label{q-formula}
\begin{figure}[!tbh]
	\centering
	\includegraphics[width=0.5\textwidth]{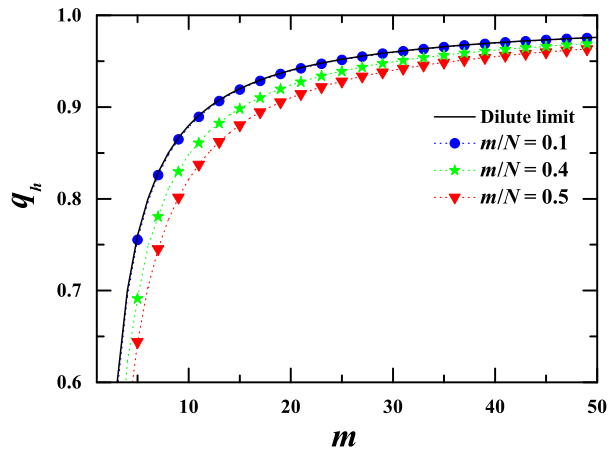}
	\caption{Variation of $q_{h}$ as a function of number of fermions for different $m/N$ values.}\label{fig.qh_1-fermi}
\end{figure}

\begin{figure}[tbh!]
	\centering
	\begin{tabular}{ll}
		\includegraphics[width=0.5\textwidth]{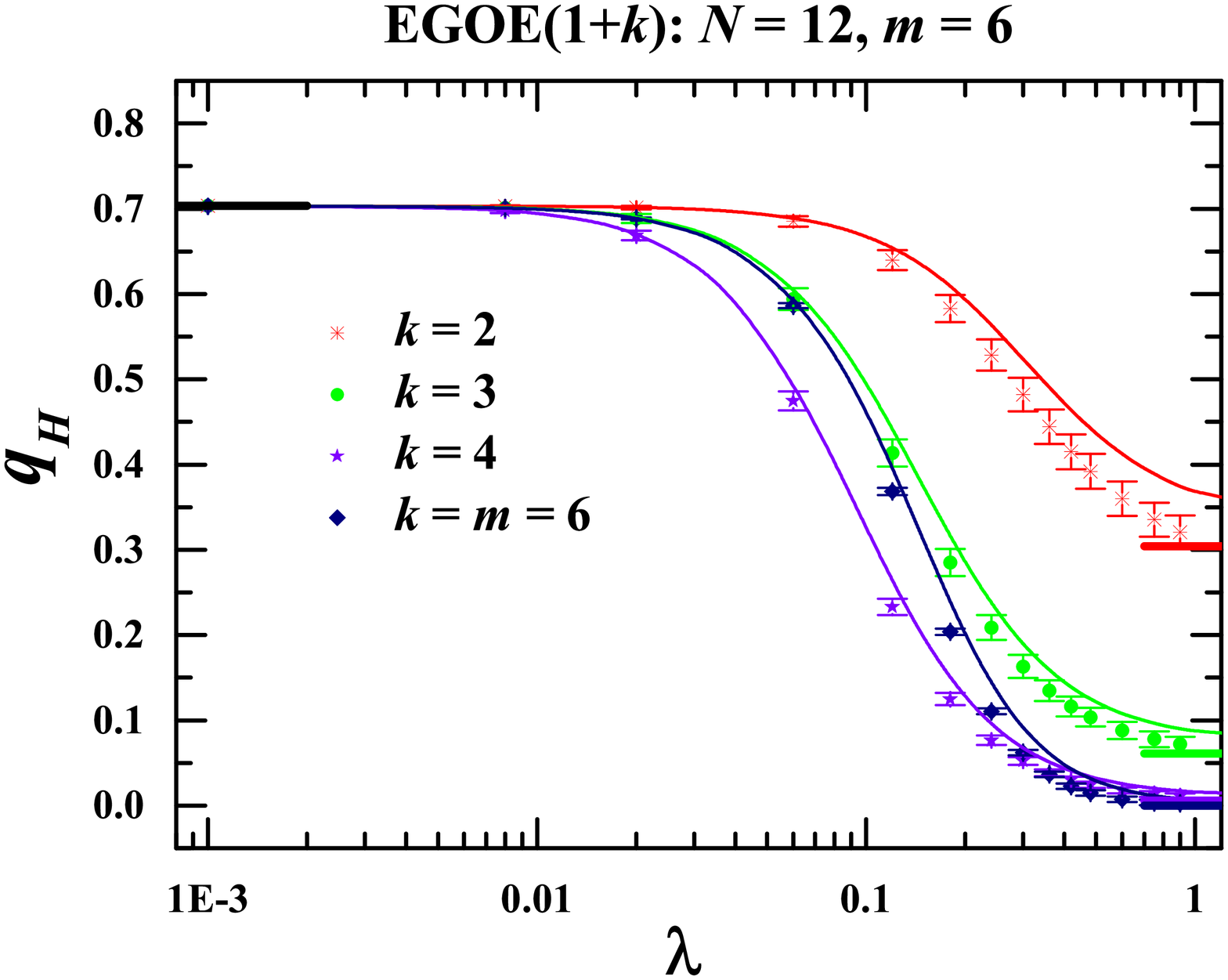}&
		\includegraphics[width=0.5\textwidth]{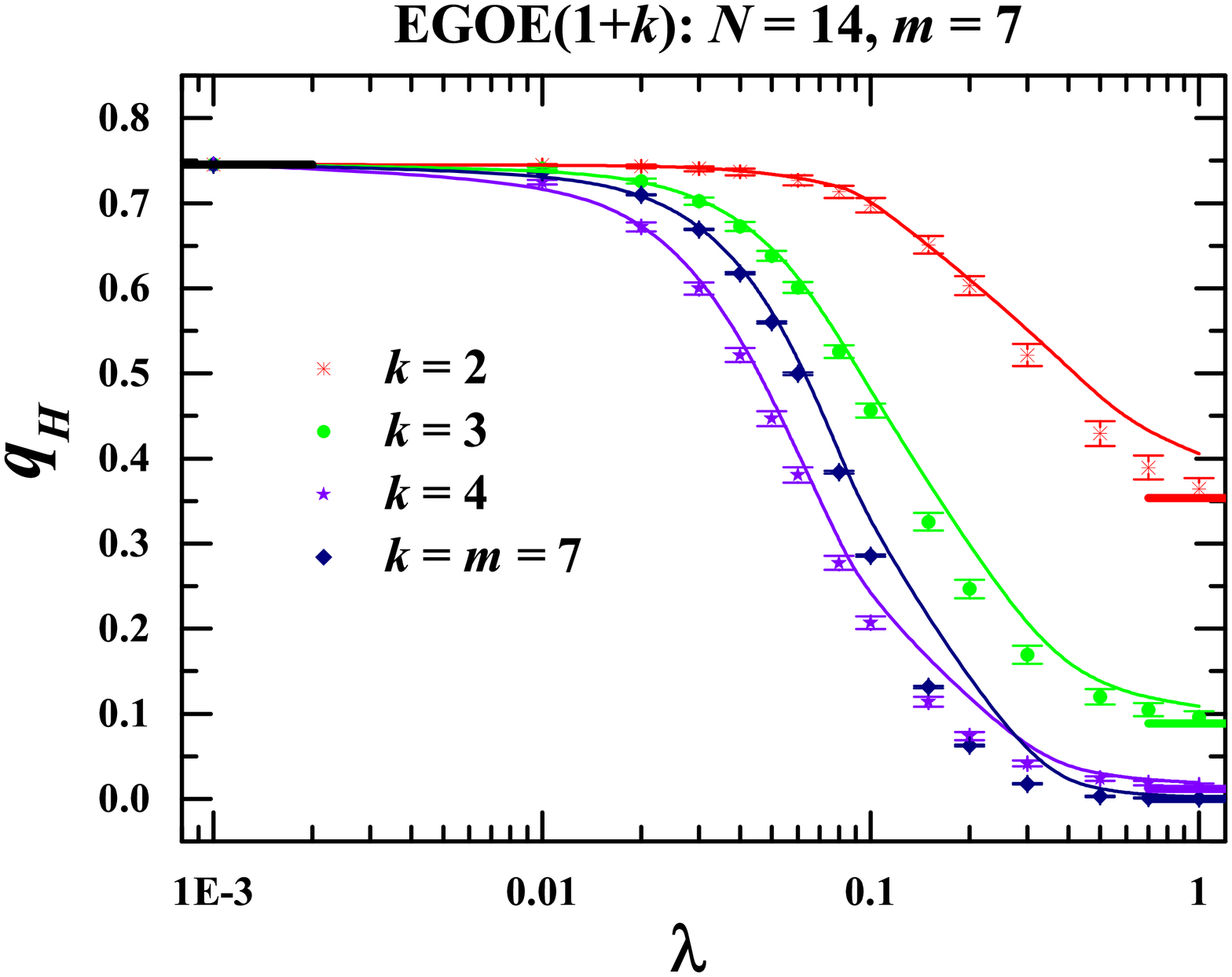}\\
		(a) & (b)
	\end{tabular}
	\caption{Variation of ensemble averaged $q$ vs. $\lambda$ for EGOE(1+$k$) ensemble for system of (a) 100 member EGOE(1+$k$) with $m = 6$ fermions in $N=12$ sp states and (b) 20 member EGOE(1+$k$) with $m = 7$ fermions in $N=14$ sp states. Results are shown for $k = 2, 3, 4$ and $k = m$. The widths of distribution of $q_H$ are shown by vertical bars.}\label{fig.qvslam-f}
\end{figure}

 {From the results shown in figure~\ref{fig.rho-q-f-1}, the ensemble averaged state density takes intermediate form between Gaussian to semi-circle as $\lambda$ and $k$ change in EGOE(1+$k$) and the univariate $q$-normal distribution $f_{N}$ can describe this transition correctly. With $\lambda=0$, the formula of $q$ for one-body part $h(1)$ only, using trace propagation formula is given by,
\be
\barr{lcl}
q_{h} +2 &=& {\langle h(1)^4 \rangle}^m \\\\
&=& \dis \frac{3 (m-1) N (N-1) (N-m-1)}{m (N-2) (N-3) (N-m)} + \dis{ \frac{N(N-1)[N^2+N-6mN+6m^2]}{m(N-m)(N-2)(N-3)} \, \frac{Y}{X^2}}
\earr
\label{eq.q-h1-f}
\ee
Here, ${\langle h(1)^4 \rangle}^m$ is the reduced fourth moment of $h(1)$ and $Y=\sum \tilde{\varepsilon_i}^4$.} Figure~\ref{fig.qh_1-fermi} represents variation of  $q_{h}$ as a function of $m$ fermions. The results are shown for various values of $m/N$ using equation~\ref{eq.q-h1-f} with fixed sp energies employed in the present study.  {From the results, one can see that $q_h$ varies from close to zero to 1 as  $m$ increases from body rank $k=1$ to higher values. This is due to fact that the form of the eigenvalue density for a $m$ particle system with $k$-body interactions is semi-circle for $k = m$ and Gaussian for $k<<m$~\cite{brody81,MF75}. In the dilute limit,  with $m \rightarrow \infty, N \rightarrow \infty, m/N \rightarrow 0$, the first term in equation~\ref{eq.q-h1-f} reduces to 3 and the second term vanishes and $q_h$ approaches to 1. The smooth curve in figure~\ref{fig.qh_1-fermi} corresponds to dilute limit. Going further, formula of $q$ parameter for complete Hamiltonian $H=h(1)+\lambda V(k)$ for fermionic systems i.e. for EGOE(1+$k$) can be given by \cite{KM2020c},
\[q_{H} + 2 = \frac{{\lan h(1)+\lambda V(k) \ran}^m}{\sigma_H^4}\;,\]
\be
q_{H} = \zeta^4\; q_{h} +  2\, \zeta^2\; (1 - \zeta^2)\; q_{hV} + (1 - \zeta^2)^2\; q_{V}.
\label{eq.q-H-f}
\ee
Here, $q_V$ represents parameter $q$ for random $k$-body interaction $H=V(k)$, with all sp energies as degenerate. The formula for $q_V$ in terms of $N$, $m$ and $k$ can be given as \cite{Manan-Ko},
\be
\barr{l}
q_{V} =
 \dis \frac{\sum_{\mu=0}^{min(k,m-k)} \Lambda^{\mu}(N,m,k)\; \Lambda^{\mu}(N,m,m-k) \;d(N:\mu)}{\binom{N}{m} \;[\Lambda^{0}(N,m,k)]^2},\\\\
 \Lambda^{\mu}(N,m,r)= \dis \binom{m-\mu}{r}\,\binom{N-m+r-\mu}{r},\;\;
 d(N:\mu)=\dis {\binom{N}{\mu}}^2-{\binom{N}{\mu-1}}^2.			
\earr \label{eq.qvk_fermi}
\ee
Similarly, $q_{hV}$ can be written as,
\be
\barr{l}
\dis q_{hV} = \dis \frac{\sum_{\mu=0}^{min(1,m-k)} \Lambda^{\mu}(N,m,k)\; \Lambda^{\mu}(N,m,m-1) \;d(N:\mu)}{\binom{N}{m} \;\Lambda^{0}(N,m,1)\; \Lambda^{0}(N,m,k)}\, .
\earr \label{eq.qv+k-f}
\ee
Figure~\ref{fig.qvslam-f} shows the results of complete variation of $q_H$ as the strength of $k$-body interaction $\lambda$ increases in EGOE(1+$k$). The $\zeta$ values are obtained using equation~\ref{eq.zeta2-f}. The formula given by equations~\ref{eq.q-H-f} is tested using the two examples considered in the present analysis: (i) a 100 member EGOE(1+$k$) ensemble with ($m=6,N=12$) and (ii) a 20 member EGOE(1+$k$) ensemble with ($m=7, N=14$). For these systems, with given $\lambda$ and body rank $k$, we construct EGOE(1+$k$) ensemble and $q_H$ is computed using the eigenvalues. The solid symbols in figure~\ref{fig.qvslam-f} represent ensemble averaged values, while the vertical bars are corresponding widths of the distributions of $q_H$ computed over members of ensemble. The continuous curves are obtained via equations~\ref{eq.zeta2-f}-\ref{eq.qv+k-f}.
The agreement between the ensemble averaged results and analytical expression is good for smaller values of $\lambda$ and $k>2$. The deviations between numerics and smooth forms for stronger interaction strength $\lambda$ with smaller $k$ are due to following reasons:(i) neglected the contribution of induced sp energies in the computation of $\zeta$ (see section~\ref{sec-zeta} for further details); (ii) The formula for $q_{hV}$, given in equation~\ref{eq.qv+k-f}, is valid for random sp energies while in the numerical results are for fixed sp energies.} In the figure~\ref{fig.qvslam-f} the horizontal marks on the left correspond to value of parameter $q_h$ and those on the right correspond to $q_V$. One can observe that when the value of $\lambda$ is very small, the ensemble averaged values of $q_H$ are found very close to  $q_{h}$ given by equation~\ref{eq.q-h1-f}. Now we gradually increase $\lambda$ and for a sufficiently large value, the ensemble averaged $q_H$ values approach the corresponding $q_{V}$ values. The formula given by equation~\ref{eq.q-H-f} gives good description of $q$ as $k$-body  interaction strength $\lambda$ varies in EGOE(1+$k$). Also, from the results shown in figure~\ref{fig.qvslam-f}, one can see that for a fixed $k$, the state density takes intermediate form between Gaussian to semi-circle as $\lambda$ changes in EGOE(1+$k$) and the transition in state density can be described using $f_{N}$ formula with values of parameter $q_H$ given by equation~\ref{eq.q-H-f}.
%%%%%%%%%%%%%%%%%%%%%%

\section{Strength functions}
\subsection{Definition and results for EGOE(1+$k$)}\label{SF}
The statistical properties related to the structure of wavefunction can be studied via so called strength functions as they carry all of the information about wavefunction structure. Since strength functions are experimentally measurable quantities they are of great importance \cite{BISZ2016}. Here, we will study strength functions to understand the structure of eigenfunctions of EGOE(1+$k$) expanded in terms of $h(1)$ basis states. Given set of basis states  $\l| \phi_\kp \ran$ for $m$-fermions distributed in $N$ sp states, the diagonal matrix elements of $H(m)$ are denoted as $\xi_{\kp}=\langle \phi_\kp | H | \phi_\kp \rangle$. Diagonalization of the full Hamiltonian $H$ gives eigenenergy $E_i$ and corresponding eigenfunction $\l| \psi_i \ran$; $H\l| \psi_i \ran= E_i \l| \psi_i \ran$. The eigenstates $\l| \psi_i \ran$ are related to $m$-particle basis states $\l| \phi_\kp \ran$ as $\l| \phi_\kp \ran = \sum_{i=1}^{d} C_{\kp}^{i}\;\l| \psi_i \ran$. Here $d$ is the dimentionality of $m$-fermion space and $C_{k}^{i}$ denotes expansion coefficients of the basis state $\l| \phi_\kp \ran$ in terms of the eigenstates $\l| \psi_i \ran$. With this we can define the strength function corresponding to a particular basis state $\l| \phi_\kp \ran$ as
\be
F_{\xi_\kp}(E) =\sum_i {|C_{\kp}^{i}|}^2 \;\delta(E - E_i)= \overline{{|\cac_{\kp}^{E}|}^2} \,I(E)\,,
\label{eq.sf}
\ee
where $I(E)$ is the state density and  $\overline{|\cac_{\kp}^{E}|^2}=\frac{1}{I (E)} \sum_{E_i\in E} {|C_{\kp}^{i}|}^2$. The definition takes into account the degeneracy in the eigen spectrum;  {Note that the last relation in equation~\ref{eq.sf} is valid only when the average is supported by chaos, not in general.}

It is well known that for $H=h(1)$, strength functions are delta functions as the basis state $\l| \phi_\kp \ran$ are identical to eigenstates $\l| \psi_i \ran$. With interaction switched on in EGOE(1+$k$), the basis state $\l| \phi_\kp \ran$  spreads into the eigenstates $\l| \psi_i \ran$ and the strength functions take the Breit–Wigner (BW) form. With further sufficient increase in $\lambda > \lambda_t$, the BW form changes to $f_{CN}$ form and ensemble averaged strength functions of many-body interacting particle systems with $k$-body interaction can be represented by smooth forms given by $f_{CN}$ \cite{PLA-2021,KM2020c}.

\begin{figure}[tbh!]
	\centering
	\begin{tabular}{cc}
		(a)&\includegraphics[width=0.7\textwidth]{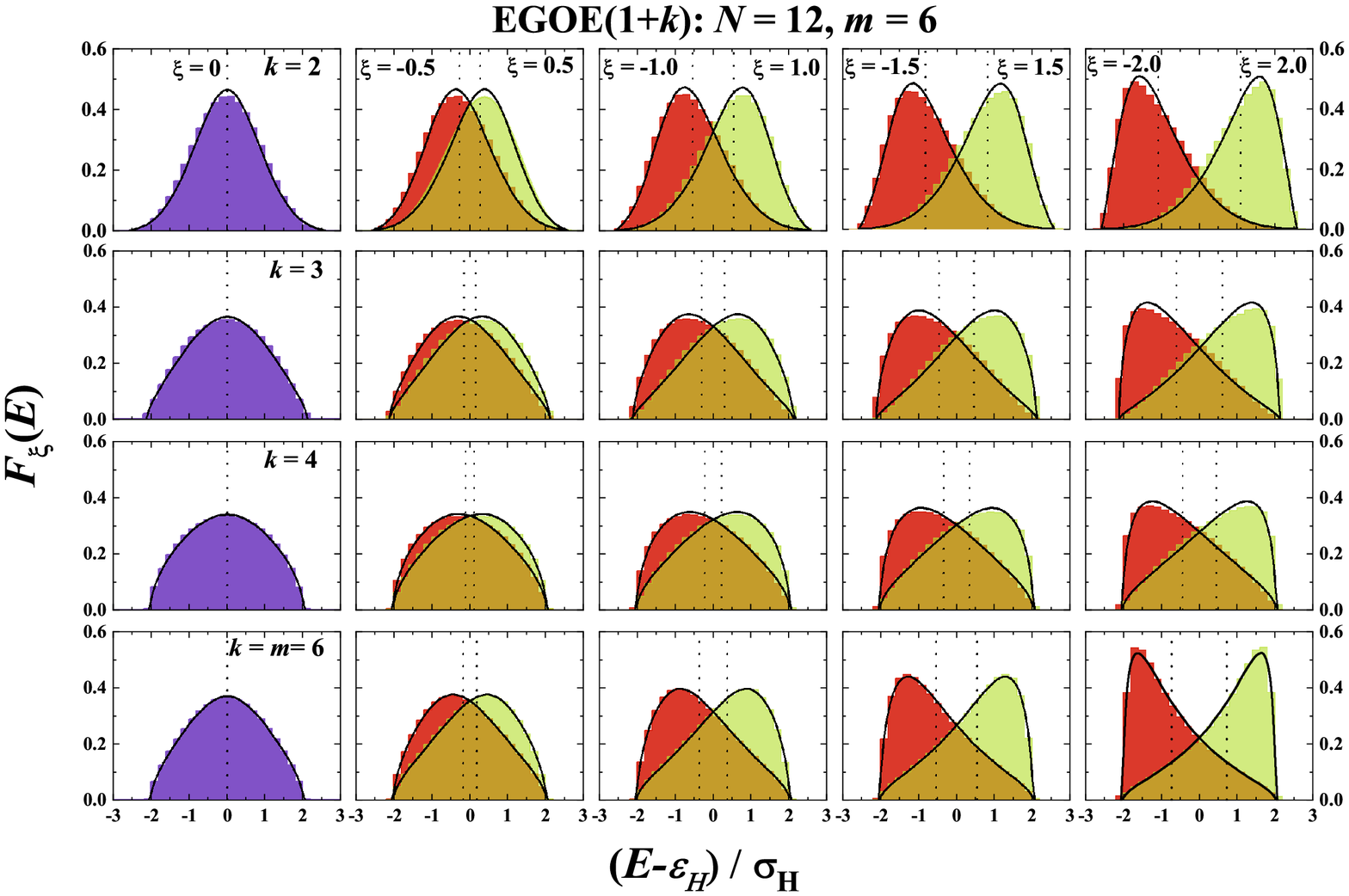}\\
		(b)&\includegraphics[width=0.7\textwidth]{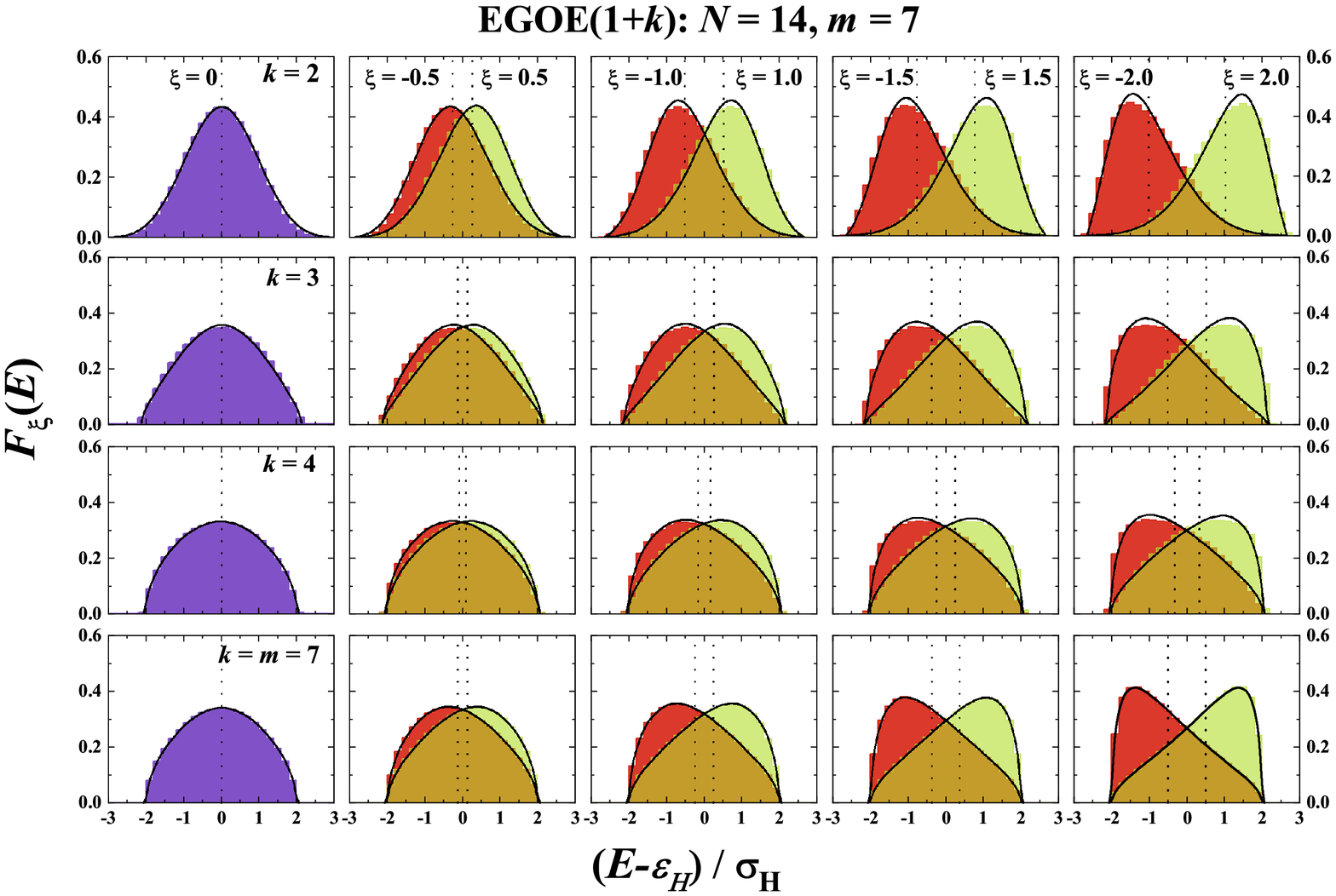}
	\end{tabular}
	\caption{Ensemble averaged strength function as a function of normalized energy $E$ for (a) a 100 member EGOE(1+$k$) ensemble with $m = 6$ and $N = 12$ and (b) a 20 member EGOE(1+$k$) ensemble with $m = 7$ and $N = 14$. Results are shown for $k$=2,3,4 and $k = m$ and  $\lambda=0.5$ is chosen for both the examples.  Histograms in extreme left panels correspond to strength functions for $\xi=0$ and histograms in the remaining panels correspond to strength functions for $\xi=\pm 0.5, \pm 1.0, \pm 1.5$ and $\pm 2.0$. In the plots $\int F_{\xi}(E) dE=1$. The continuous curves are due to fitting with $f_{CN}(\hat{E}|\hat{\xi};\zeta,q=q_H)$ given by equation~\ref{eq:biv-cqn} with corresponding ensemble averaged $q_H$ and $\zeta$ values. The vertical dotted lines, in each panel, correspond to centroid, $\varepsilon_F=\zeta\,\xi$. See text for more details.}
	\label{fig.fke}
\end{figure}

%%%%%%%%%%%%%%%%%
In order to compute the ensemble averaged strength function $F_\xi(E)$ in EGOE(1+$k$), with $\lambda$ sufficiently large where $f_{CN}$ form is applicable, we adopted the procedure used in \cite{PLA-2021}. In the ensemble averaged calculations, the $E_i$ and $\xi_\kp$ spectra are scaled to have zero centroid and unit width for each member of the ensemble using the eigenvalue distribution and  $\kp$-energies distribution, respectively. Therefore, $E \rightarrow \hat{E}$ and  $\xi/\sigma_H=\zeta\,\hat{\xi}$. Now, with this scaling the method of construction of strength function used in the previous studies can apply \cite{Ch-04,CK2017,PLA-2021}. Figure~\ref{fig.fke} shows the ensemble averaged strength function results obtained for
two EGOE($1+k$) examples with (i) $m = 6$ fermions in $N=12$ sp states and (ii) $m = 7$ fermions in $N=14$ sp states. Here $\lambda$ is chosen equal to 0.5, so that the system is in thermalization region \cite{Manan-Ko}. The ensemble averaged $F_{\xi}(E)$ results are shown for  $\hat{\xi} = 0.0, \pm 0.5, \pm 1.0, \pm 1.5$ and $\pm 2.0$ using body rank $k = 2, 3, 4$ and $m$ for both the examples. All these
strength function histograms $F_\xi(E)$ are fitted  with $f_{CN}(\hat{E}|\hat{\xi};\zeta,q=q_H)$.
For each $k$, the smooth curves in figure~\ref{fig.fke} are obtained with $f_{CN}$ using corresponding ensemble averaged values of $\zeta$ and $q_H$. It is clearly seen from the results shown in the figure that the ensemble averaged histograms are in very good agreement with the smooth forms obtained using $f_{CN}$.
The vertical dotted lines in each panel of figure~\ref{fig.fke}, represent the centroid $\varepsilon_F$ and it is equal to $\zeta\,\xi$.  Following conclusions can be derived from the results shown in the figure~\ref{fig.fke}: (i) One clearly sees that for sufficiently large $\lambda$, the smooth part of strength functions $F_\xi(E)$ are very well represented by $f_{CN}$ and they make a transition from Gaussian form to semi-circle form, as the body rank $k$ in EGOE(1+$k$) changes from 2 to $m$. (ii)  $F_\xi(E)$ results for $\hat{\xi}=0$ are symmetric (extreme left panels in figure~\ref{fig.fke}). (iii) Now for $\hat{\xi} \neq 0$, away from the center of the spectrum, $F_\xi(E)$ results are asymmetrical about $\hat{E}$ as demonstrated earlier \cite{CK2017,PLA-2021}. (iv) The centroid of $F_\xi(E)$ changes with $k$ for $\hat{\xi} \neq 0$. (v) $F_\xi(E)$ are more skewed in the positive direction for negative $\hat{\xi}$ values and are more skewed in the negative direction for positive $\hat{\xi}$ values. From these results and also obtained in \cite{PLA-2021,KM2020c}, one can clearly conclude that the strength functions of many-particle systems with random $k$-body interactions, follow the conditional $q$-normal distribution $f_{CN}$.
 { It is important to note that for $\lambda = 0$, i.e. with one-body interaction alone, the basis states themselves are the eigenstates and hence $F_\xi(E)$ then corresponds to the $\delta$ function for all $k$. With increase in $\lambda$, the shape of $F_\xi(E)$ changes from the Breit-Wigner (BW) form.	
With further increase in $\lambda$, the shape of $F_\xi(E)$ changes to Gaussian or semi-circle or intermediate to Gaussian and semi-circle depending upon the values of $\lambda$ and $k$. Therefore, one can also observe the BW to Gaussian to semi-circle transition in $F_\xi(E)$ by changing both $\lambda$ and $k$. The $f_{CN}$ can not give BW form for $F_\xi(E)$ in any limit and it is possible to study BW to semi-circle transition in $F_\xi(E)$ using hybrid expression given in \cite{zelev-95}.
}

Very recently, it is shown that for EGOE(1+$k$) with sufficiently large $\lambda$, the first four moments of $F_\xi(E)$ are very close to that of $f_{CN}$, analytically \cite{KM2020c} as well as numerically \cite{PLA-2021}.
In the next section, we will further analyze strength functions of EGOE(1+$k$) by computing their first four moments and to identify the region of thermalization by comparing them with corresponding smooth expressions given by $f_{CN}$.
%%%%%%%%%%%%%%%%%%%%

\subsection{Lower order moments of Strength functions }
 \label{zeta}

 \subsubsection{Centroid $\varepsilon_F$}
\begin{figure}[tbh!]
	\centering
	\begin{tabular}{cc}
		{\bf $k=2$}&{\bf $k=3$}\\
		\includegraphics[width=0.5\textwidth]{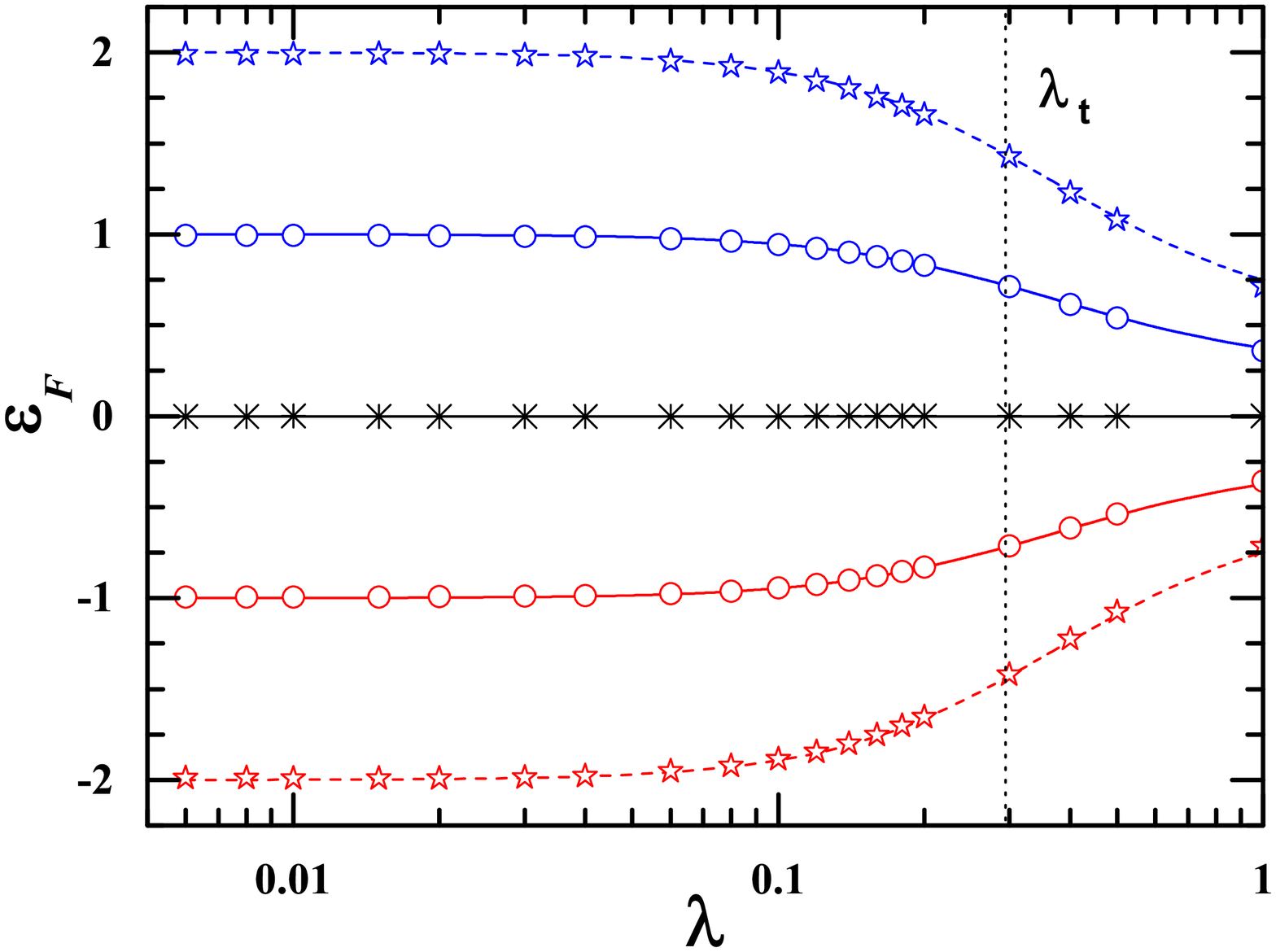}&\includegraphics[width=0.5\textwidth]{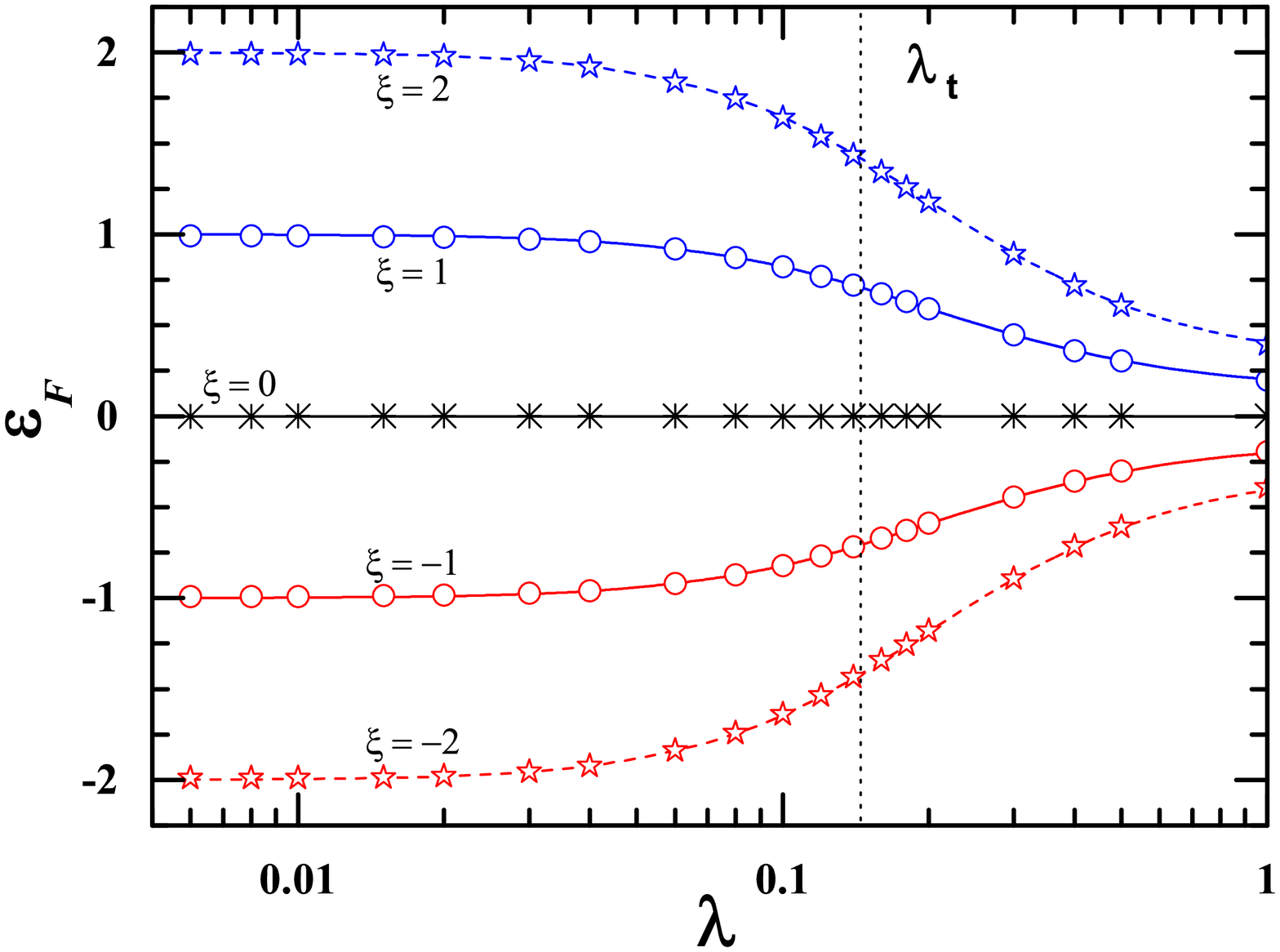}\\
		(a)&(b)
	\end{tabular}
	\caption{Centroid of the strength functions vs. $\lambda$ for the example of 100 member EGOE(1+$k$) with $m$=6 fermions in $N$=12 sp states for (a) $k = 2$ and (b) $k= 3$. The results are shown for various values of $\xi$. For more details refer the text.}\label{fig.ef-lt}
\end{figure}

 From the results of previous section, it is clear that asymmetry in strength function with respect to $\hat{E}$ and $\hat{\xi}$ increases and also
 the centroid of the strength function moves with $\hat{\xi}$. The centroid of $f_{CN}$ is given by $\varepsilon_F= \zeta\,\hat{\xi}$ and it is compared with the ensemble averaged centroid of $F_\kp(E)$ computed for a 100 member EGOE(1+$k$) with $(m,N) = (6,12)$. Figure~\ref{fig.ef-lt} represents the variation of $\varepsilon_F$ with $\lambda$ for $k=2$ and 3. The ensemble averaged results (symbols) are compared with expression given by $f_{CN}$ (smooth curves) and the match between them is exact. From the results, it is clear that the peak values of $F_\kp(E)$ change with $\hat{\xi}$ and the rate of change of $\varepsilon_F$ with respect to $\hat{\xi}$ is given equal to the correlation coefficient $\zeta$.

%%%%%%%%%%%%%%
%%%%%%%%%%%%%%%%%
 \subsubsection{Width $\sigma_F$ and thermalization marker $\lambda_t$} \label{section5b}

\begin{figure}[tbh!]
	\centering
	\includegraphics[width=0.55\textwidth]{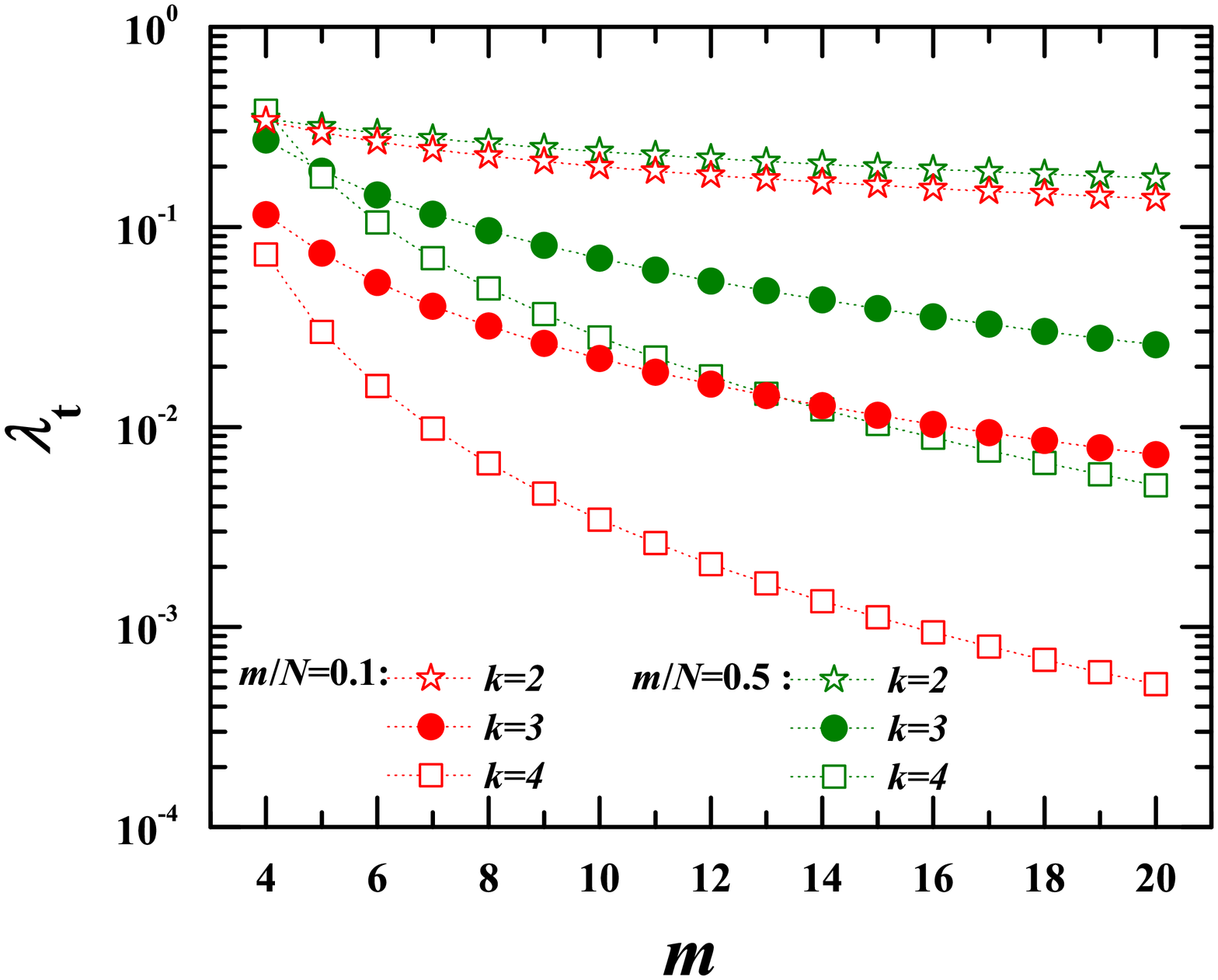}
	\caption{Marker $\lambda_t$ vs. $m$ for fermion systems. Results are obtained using equation~\ref{eq.lt} for $k$=2,3 and 4 and for various values of $m/N$.}
	\label{fig-lt}	
\end{figure}

The expression for the normalized width of the strength functions $\sigma_F$ is exactly same as that of $f_{CN}$ and it is $\sigma_F= \sqrt{1-\zeta^2}$. The width $\sigma_F$ (independent of $\xi$) of the strength functions is scaled by the spectrum width $\sigma_H$ and $\zeta$ is the correlation coefficient which determines the thermodynamic chaos marker \cite{PLA-2021}. Therefore, the variation of $\sigma_F$ can be discussed in terms of $\zeta^2$ results given in Section~\ref{sec-zeta}. The thermodynamic region is the region where different quantities defining the eigenstate properties like entropy, strength functions, temperature, etc. give the same values irrespective of the defining basis. This means that once the interaction strength $\lambda$ between the particles goes beyond $\lambda_t$, the quantum system thermalizes and then we can define quantities like temperature, entropy, etc. of the system.
Very recently, analytical expression of $\lambda_t$ is obtained for interacting dense boson systems in \cite{PLA-2021}.  {The critical value of the interaction strength is obtained extending the concept of duality to EGOE(1+$k$). As first discussed in ~\cite{Jac2002}, at duality point $\lambda=\lambda_t$, the quantities defining eigenstate properties like strength functions, entropies, temperature, etc. give the same values irrespective of the defining basis. With $H=h(1)+\lambda V(k)$, we have two choices of basis: one defined by one-body $h$ and other by infinite interaction $V$. The variance of one-body part can be given  as $\sigma_h^2=a^2\,\Delta^2$ and that of interaction part $\sigma_V^2=b^2\,\lambda^2$. Now, for the strength functions expanded in one-body basis,  the correlation coefficient is by $\zeta_h=\sqrt{a^2\,\Delta^2/(a^2\,\Delta^2+b^2\,\lambda^2)}$. Similarly, for the strength functions expanded in interaction basis, the correlation coefficient is by $\zeta_V=\sqrt{b^2\,\lambda^2/(a^2\,\Delta^2+b^2\,\lambda^2)}$. At duality point $\lambda=\lambda_t$, it can be seen that the spreading produced by one-body part and $k$-body part are equal ($\zeta_V=\zeta_h$) giving condition that $\zeta^2 = 0.5$ \cite{AGK2004}}. Putting $\zeta^2 = 0.5$ in equation~\ref{eq.zeta2-f} and solving it for $\lambda$, ($m$, $N$, $k$) dependence of marker $\lambda_t$ is obtained as,
\be
\lambda_t=\sqrt{\frac{m (N-m)\; X}{ N(N-1)\Lambda^0(N,m,k)(1-3\;\binom{N}{k}^{-1})}}\;\;.
\label{eq.lt}
\ee
Using above equation, we have $\lambda_t(k = 2) = 0.2944$, $\lambda_t(k = 3) = 0.1448$ and $\lambda_t(k = 4) = 0.1053$ for $m=6,N=12$). Similarly, with ($m=7,N=14$), $\lambda_t(k = 2) = 0.2764$, $\lambda_t(k = 3) = 0.1158$ and $\lambda_t(k = 4) = 0.0696$. $\lambda_t$ results for $k=2$ are consistent with theoretical estimates obtained in \cite{AGK2004}. The results of the variation of marker $\lambda_t$ (which defines the thermodynamic region) as a function of $m$ are presented in figure~\ref{fig-lt}. Results are shown for $m/N$ values 0.1 and 0.5 for $k$ = 2, 3 and 4. It can be clearly seen that  value of $\lambda_t$ decreases as the body rank of interaction $k$ among the particles increases i.e. the thermalization sets in faster in the system with increase in rank of interaction $k$.  {Further, in the results of $\zeta^2$ given in figure~\ref{fig-zeta2-f}, the vertical dash-lines  indicate the positions of $\lambda_t$ for given $k$ obtained using equation~\ref{eq.lt}. It
can be clearly seen that $\zeta^2 = 0.5$ gives the thermalization point $\lambda_t$ correctly for given body rank of the interaction $k$.}

 {Now in the next section, we will analyze the ensemble averaged results of skewness $\gamma_1$ and excess $\gamma_2$ of $F_\xi(E)$ as a function of $\lambda$ with that of $f_{CN}$ and show that they can be used to identify the thermodynamic region $\lambda_t$.}
%%%%%%%%%%%%%%%%%%%%%%
%%%%%%%%%%%%%%%%%%

\subsubsection{Skewness $\gamma_1$ and Excess $\gamma_2$}
\begin{figure}[tbh!]
	\centering
	\begin{tabular}{cc}
		{\bf $k=2$}&{\bf $k=3$}\\
		\includegraphics[width=0.5\textwidth]{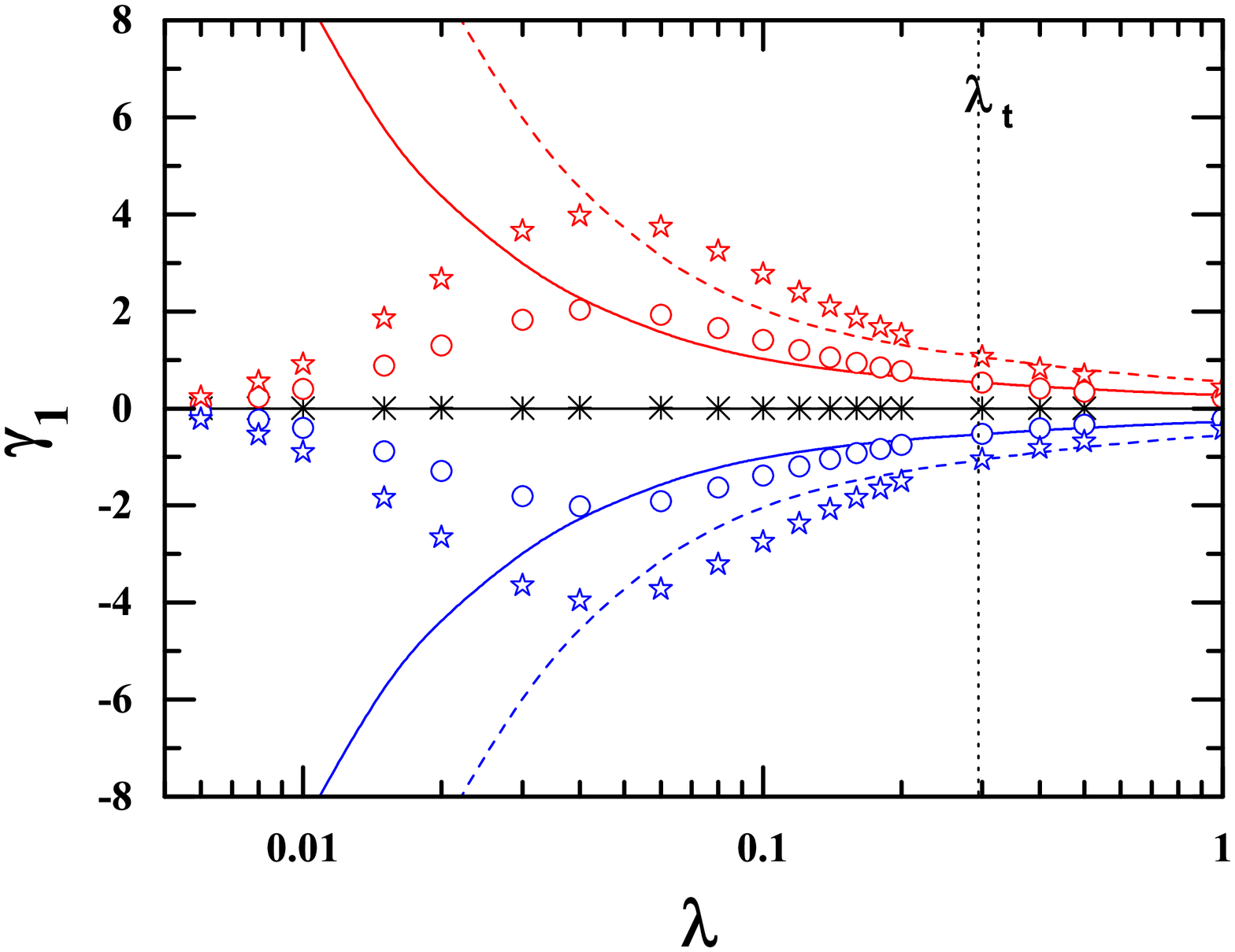}&\includegraphics[width=0.5\textwidth]{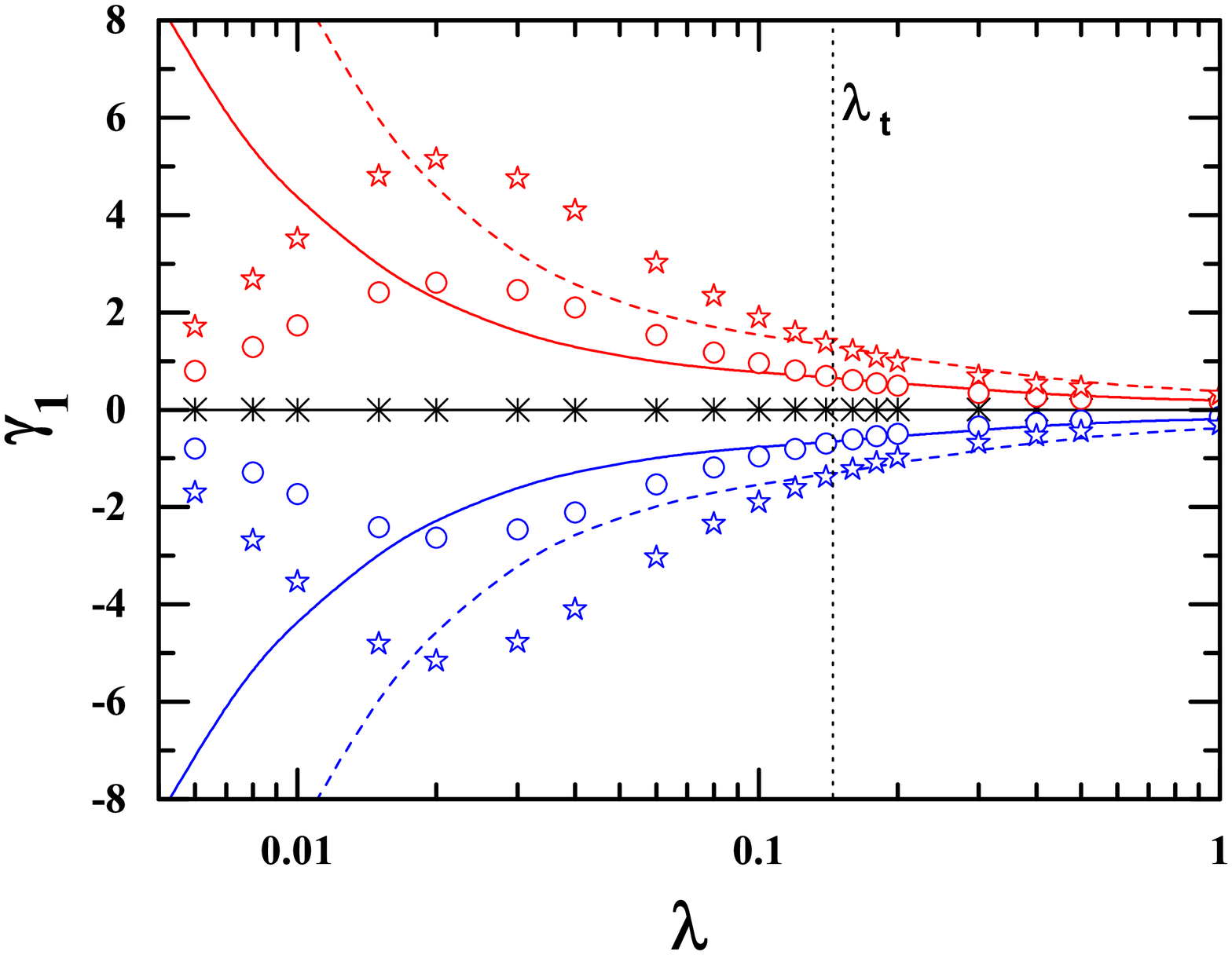}\\
		\includegraphics[width=0.5\textwidth]{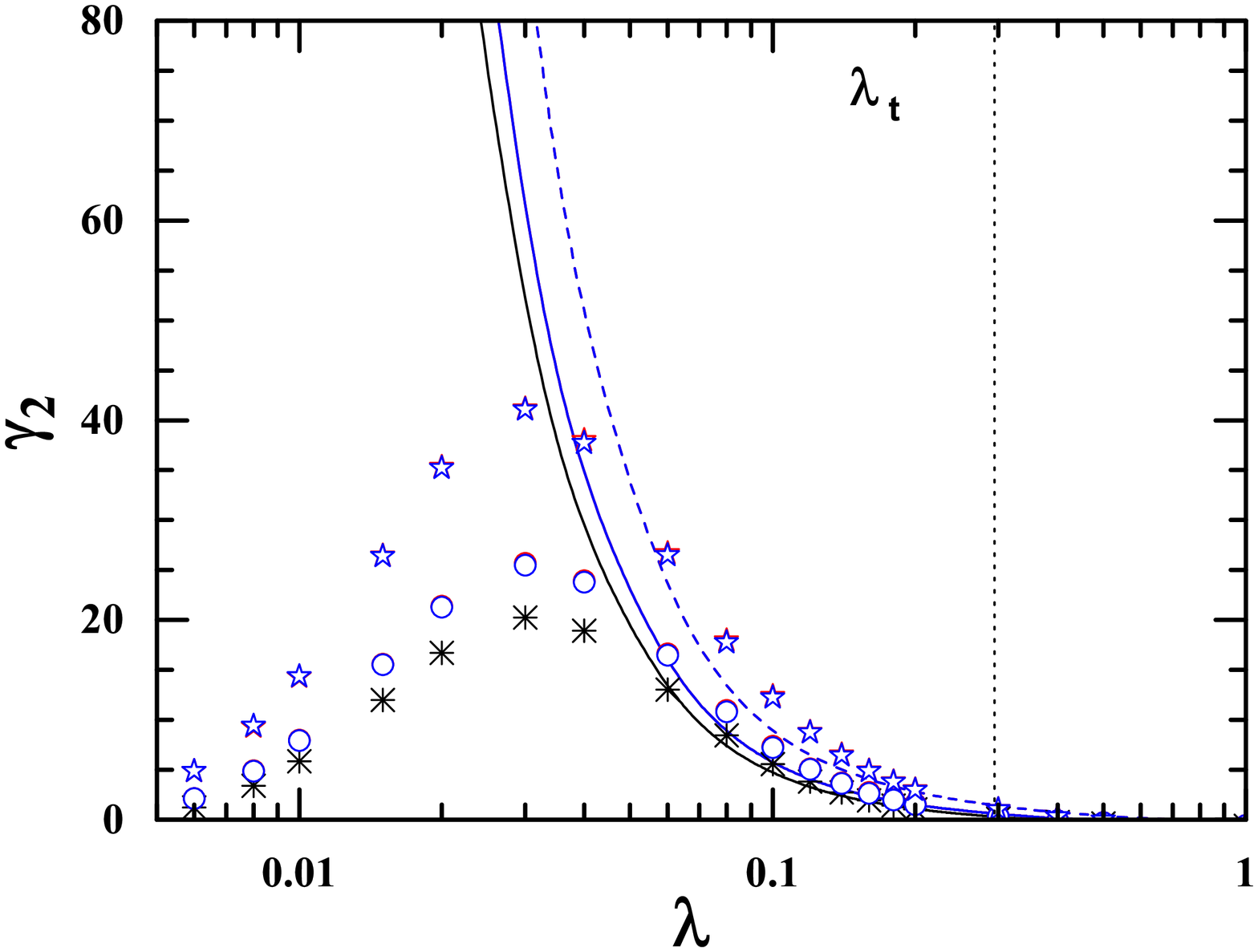}&\includegraphics[width=0.5\textwidth]{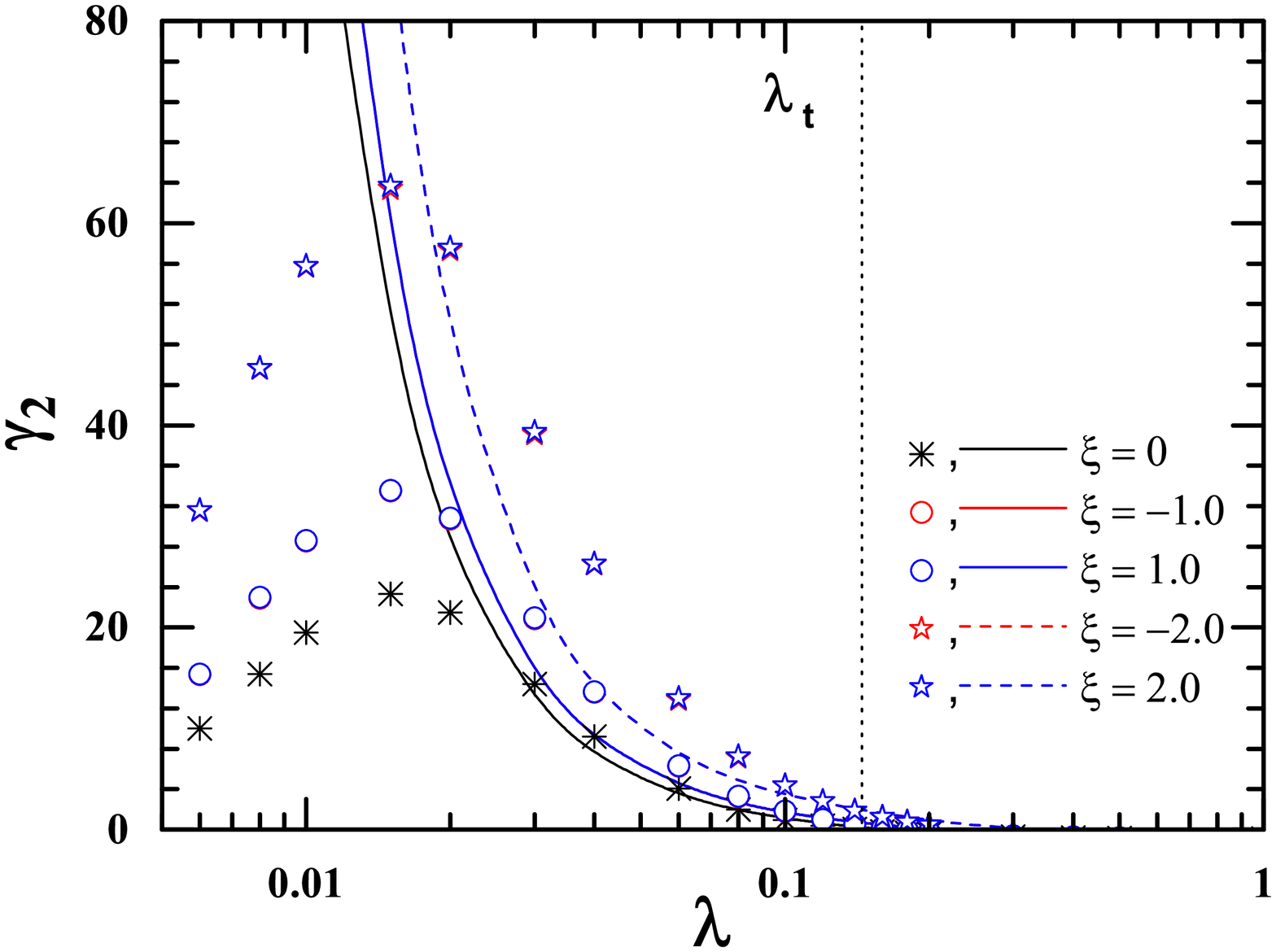}\\
		(a)&(b)
	\end{tabular}
	\caption{Variation of $\gamma_1$ and $\gamma_2$ of strength functions as a function of $\lambda$. The results are shown for various values of $\xi$ for the example of 100 member EGOE(1+$k$) with $m$=6 fermions in $N$=12 sp states for (a) $k = 2$ and (b) $k= 3$. The vertical dotted
		lines in each panel indicate position of $\lambda_t$. For details refer the text.}\label{fig.mom-lt}
\end{figure}

%Now, we will compare the ensemble averaged results of skewness $\gamma_1$ and excess $\gamma_2$ of the strength function  $F_\xi(E)$ with that of $f_{CN}$.
The simple analytical expressions for $\gamma_1$ and  $\gamma_2$ of $f_{CN}$ can be given as  \cite{KM2020c},
\be
\barr{rcl}
\gamma_1(\hat{\xi}) &=& - \dis \frac{\varepsilon_F\,(1-q)}{\sigma_F},\\\gamma_2 \dis (\hat{\xi}) &=& (q-1) + \gamma_1^2\, \l(1+\frac{(1+q)}{(1-q){\hat{\xi}}^2}\r)\;,
\earr
\label{eq:mom}
\ee
We have computed ensemble averaged $\gamma_1$ and $\gamma_2$ of $F_{\xi}(E)$ for a 100 member EGOE(1+$k$) with $m=6$ fermions in $N=12$ sp states as a function of $\lambda$, for body rank of interaction $k = 2$ and 3, and  the results are shown in figure~\ref{fig.mom-lt} for various values of $\hat{\xi}=0$, $\pm 1.0$, $\pm 2.0$. The solid symbols represent EGOE(1+$k$) results while the smooth forms are obtained using the formulas given in equation~\ref{eq:mom}. From the results shown in the figure, it is clear that $\gamma_1$ is non zero for $\hat{\xi} \ne 0$, $\gamma_1$ is positive for $\hat{\xi} <0$ and  negative for $\hat{\xi} >0$. This feature is clearly visible in the strength function results presented in previous section. Further, $\gamma_2$ is always positive for $\pm \hat{\xi}$. In the thermodynamic region with $\zeta^2 \le 0.5$, $\gamma_2 \approx (q-1)$ as suggested in \cite{PLA-2021}. For smaller value of $\lambda$, the deviation between numerical results  and the corresponding smooth forms of $\gamma_1$ and $\gamma_2$, are very large indicating $f_{CN}$ forms can not be applied to represent strength functions $F_{\xi}(E)$ at this stage. Further, for sufficiently large $\lambda$($ \geq \lambda_t$) where $f_{CN}$ form is applicable to $F_{\xi}(E)$, the deviation between ensemble averaged results and smooth forms for $\gamma_1$ and $\gamma_2$ decreases to zero giving onset of thermalization marker $\lambda_t$. In figure~\ref{fig.mom-lt}, $\lambda_t$ is shown by vertical dotted lines for each $k$ results. The $\lambda_t$ values obtained from moments are in good agreement with those obtained from analytical results using equation \ref{eq.lt}. Hence, the onset of thermalization in quantum many-fermion systems with $k$-body interactions can also be located using the moments of $F_{\xi}(E)$.

\section{Inverse Participation Ratio (IPR)}\label{NPC}

\begin{figure}[tbh!]
	\centering
	\begin{tabular}{cc}
		\includegraphics[width=0.5\textwidth]{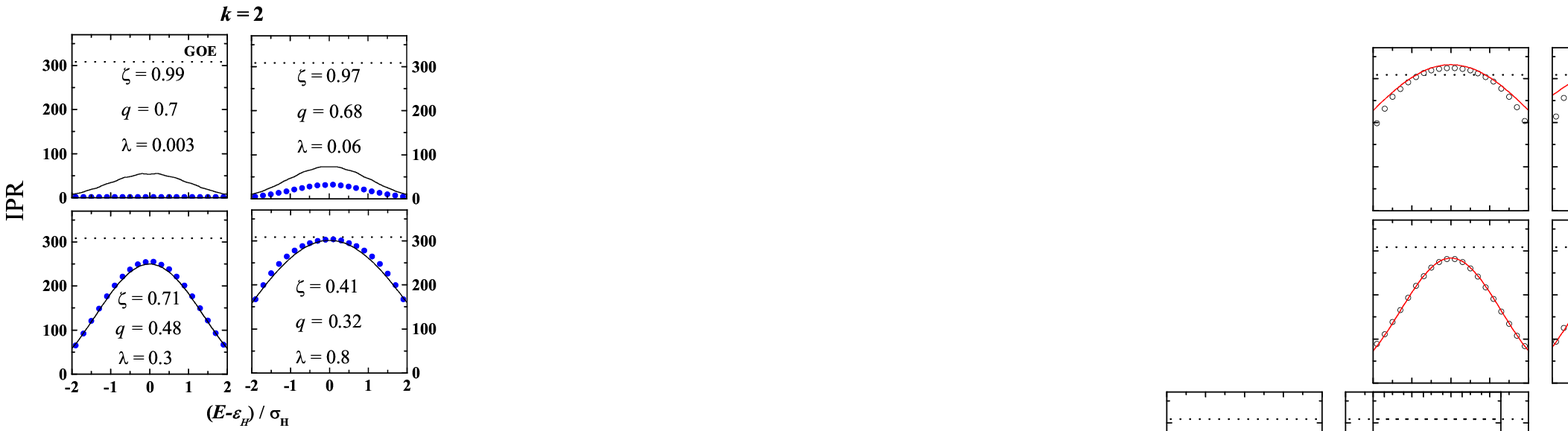}&\includegraphics[width=0.5\textwidth]{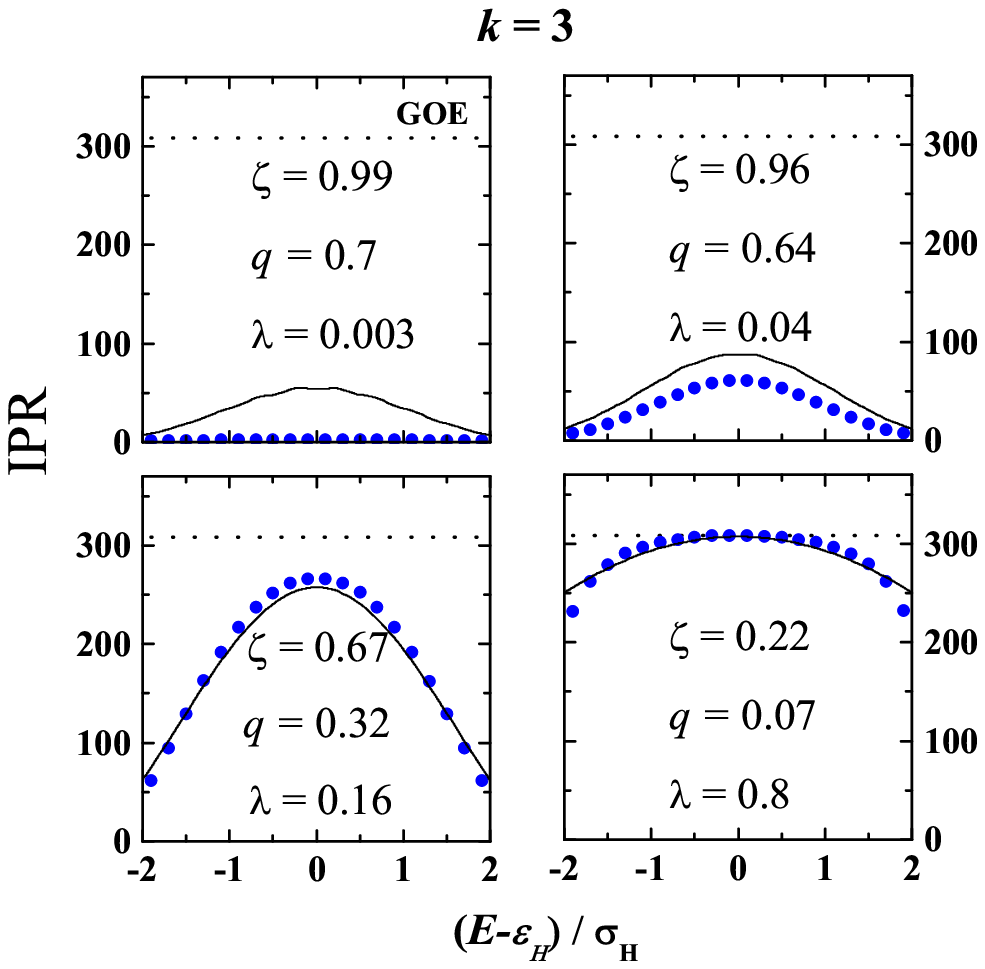}\\
		\includegraphics[width=0.5\textwidth]{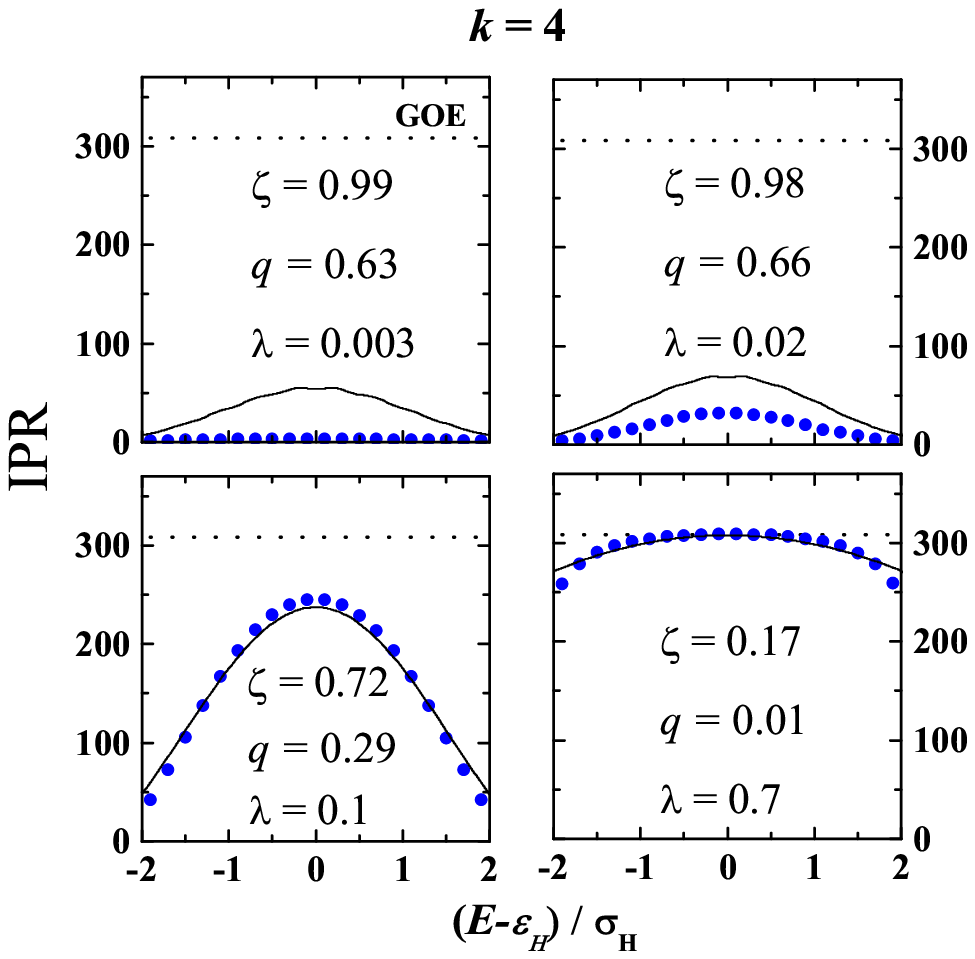}&\includegraphics[width=0.5\textwidth]{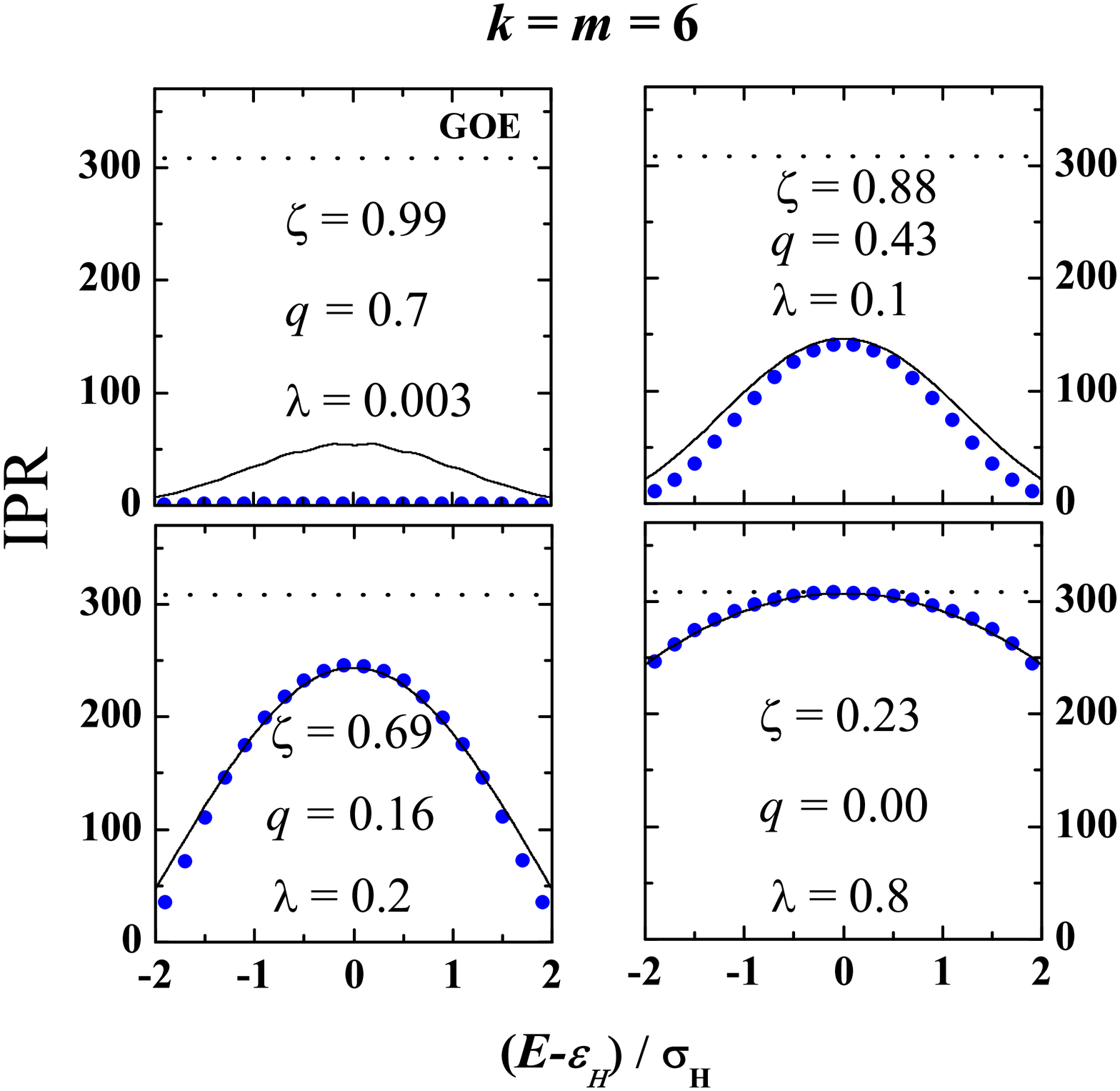}
	\end{tabular}
	\caption{Ensemble averaged IPR as a function of normalized energy $E$ for EGOE(1+$k$) with $m=6$ fermions distributed in $N=12$ sp states for body rank $k=2, 3, 4$ and $k = m = 6$. Here an ensemble of 100 members is considered. Results are shown for various values of $\lambda$. Here blue solid circles correspond to numerical ensemble averaged EGOE(1+$k$) results and the continuous black curves correspond to theoretical curves obtained via equation~\ref{eq.npc}. GOE estimate is shown by dotted lines in each graph.}
	\label{fig-npc-f} 	
\end{figure}

 {IPR (also known as NPC) is a useful quantity to measure the degree of delocalization of a given eigenstate within a specific basis. It gives essentially the number of basis states $|\phi_\kp \rangle$ that make up the eigenstate $|\psi_i \rangle$ with energy $E$. For an eigenstate $|\psi_i \rangle$ spread over the basis states $|\phi_\kp \rangle$, with energies $\xi_\kp = \langle \phi_\kp|H|\phi_\kp \rangle$, it is defined as,
\be
\mbox{IPR}(E) = \left\{ {\dis \sum\limits_\kp {\left| {C_{\kp}^{i} } \right|^4 } } \right\}^{ - 1}.
\ee
For the case of full localization, the $|\psi_i \rangle$ state is one of the basis states, so only one $C_{\kp}^{i} = 1$ giving IPR close to 1. On the other hand, if the state $|\psi_i \rangle$ is equally distributed among all basis states giving GOE value of IPR =$d/3$ with $d$ is the dimensionality of the $m$-particle space. Recently, a simple two-parameter expression for IPR$(E)$, applicable to one-body plus an embedded GOE of $k$-body interaction for interacting many-particle systems, is derived in \cite{PLA-2021}.} This expression is valid in the chaotic domain ($\lambda \le \lambda_t$) where the level fluctuations follow GOE \cite{ndc} and the form of the strength functions is very well represented by $f_{CN}$. This expression for IPR$(E)$ is given as,
\be
\mbox{IPR}^{\zeta,q}(E) = \dis\frac{d}{3} \dis \left\{ \sum_{n=0}^{\infty} \frac{\zeta^{2n}}{[n]_q!}\, [\bH_n(\hat{E}|q)]^2 \right\}^{-1} ,
\label{eq.npc}
\ee
The parameters $\zeta$ and $q$ are described in sections~\ref{sec-zeta} and \ref{q-formula}, respectively.  {The equation~\ref{eq.npc} can be applicable to fermionic EGOE(1+$k$) as it depends only on the parameters, $zeta$ and $q$, defining the shape of $f_N$ and $f_{CN}$.} The equation~\ref{eq.npc} is tested with ensemble averaged EGOE(1+$k$) results for fermion systems for various values of $k$. Ensemble averaged IPR results are presented in figure~\ref{fig-npc-f}, by considering a 100 member EGOE(1+$k$) ensemble with $m=6$ fermions in $N = 12$ sp states example for different values of $\lambda$ for body rank $k = 2, 3, 4$ and $k = m = 6$. The blue solid circles represent the ensemble averaged IPR values. The smooth curves are due to  the theoretical expression given by equation~\ref{eq.npc} and the ensemble averaged $\zeta$ and $q$ values used for this analysis are given in the corresponding figures. One can clearly see the localization to delocalization transition with increase in $\lambda$ for fixed $k$ from these results. When the value of $\lambda$ is very small, i.e. when the one-body part of the interaction dominates over the $k$-body interaction, the IPR values are close to one and a large deviation between the numerical results and the theoretical curves is observed. This indicates the wavefunctions are completely localized in this region. Now as $\lambda$ is slightly increased, the deviation between the numerical results and the theoretical curves is still observed, however it is reduced compared to the previous case as the chaos has not yet set in the system. Now when $\lambda$ is increased to a sufficiently high value ($\lambda = \lambda_t$), one can see that in the bulk of the spectrum there is a good match between ensemble averaged results and theoretical curves whereas small deviations are observed near the spectrum tails. In this region the system is in chaotic domain corresponding to the thermalization region given by $\zeta^2 \sim 1/2$ \cite{Ch-PLA} and the smooth variation of strength functions $F_\xi(E)$ are well represented by $f_{CN}$ and IPR reaches the theoretical chaotic domain estimate value for all $k$. It can be clearly seen from the results that for $\lambda \geq \lambda_t$, the match of the theoretical expression and the ensemble averaged numerical results, is very good for all $k$; indicating onset of thermalization and the wavefunctions are delocalized in this region.
%%%%%%%%%%%%%%%%%%

\section{Conclusions}\label{conclusions}
In the present paper, we have studied the conditions of thermalization in fermion systems with EGOE(1+$k$) using the  measures based on the structure of eigenfunctions, the shape of strength functions and IPR, as a function of the interaction strength $\lambda$ and for various values of body rank of the interaction $k$.
%%%%%%%%%%%%%%
It is shown that $q$-normal distribution $f_{N}$ defining $q$-Hermite polynomials can describe the smooth part of the state density in interacting fermion systems in dilute limit just like their bosonic counter part in dense limit \cite{PLA-2021}. An  analytical expression is derived for the $q_h$-parameter for fixed one body part $h(1)$ and it is shown that in the dilute limit $q_h \rightarrow 1$. Also a complete description of $q_H$-parameter is obtained as a function of the strength of interaction $\lambda$ for various values of body rank $k$ in fermion systems with EGOE(1+$k$). Further, we have shown that the strength function makes transition from
Gaussian form to semi-circle form as $k$ changes from 2 to $m$ with sufficiently large $\lambda$ value and it can be represented by conditional $q$-normal density described by $f_{CN}(\hat{E}|\hat{\xi}=0,\zeta,q_H)$ at the center of the spectrum as described in \cite{KM2020c}.
For $\hat{\xi} \neq 0$, i.e. away from the spectrum, it is clearly seen that the strength functions can also be represented correctly by     $f_{CN}(\hat{E}|\hat{\xi},\zeta,q_H)$ with proper scaling. Similar results were also obtained for bosonic systems in the dense limit \cite{PLA-2021}. From the results one can see that for $\hat{\xi} \neq 0$ as one moves away from the center of the spectrum, the strength functions become asymmetric and this asymmetry is specifically described by $f_{CN}$. Going further, the first four moments of the strength function are studied and utilized to identify the region of thermalization. The numerically computed values of the correlation coefficient $\zeta$, which is related to the width of the strength functions ($\sigma_F$), match with the trace propagation results with varying $\lambda$ for all $k$ values and in the dilute limit $\zeta \rightarrow 0$. Further, we obtain marker $\lambda_t$ in terms of $m$, $N$ and $k$ and it is shown that $\lambda_t$ decreases as the body rank $k$ of interaction increases indicating faster on set of thermalization with $k$. Further, $\lambda_t$ obtained via moments of strength functions is in good agreement with the analytical results. Going further, we have studied chaos measure IPR as a function of $\lambda$ for different values of $k$ and found that the ensemble averaged results are in very good agreement with the two-parameter ($q$ and $\zeta$) formula derived in \cite{PLA-2021} in the thermalization region ($\lambda \geq \lambda_t$). Results of the present investigation along with obtained in  \cite{PLA-2021,Manan-Ko,KM2020,KM2020c,ndc}, it is clear that the $q$-Hermite polynomials play a very significant role in the analysis of many-body quantum systems interacting via $k$-body interaction.

\section*{Acknowledgements}
Thanks are due to V. K. B. Kota for many useful discussions. This work is partially supported by Department of Science and Technology (DST), Government of India [Project No.: EMR/2016/001327]. NDC acknowledges financial support from University
supported research project [grant No: GCU/RCC/2021-22/20-32/508]. PR acknowledges support from ICTS for participating in the online program - Bangalore School on Statistical Physics XII (Code: ICTS/bssp2021/6).

\end{document}